\documentclass{aa}  

\usepackage[colorlinks=true,     linkcolor=blue, citecolor=blue, filecolor=blue, urlcolor=blue]{hyperref}
\usepackage{lineno}
\usepackage{multirow}
\RequirePackage{linenoaa}

\linenumbers

\usepackage{graphicx}

\usepackage{txfonts}

\usepackage{soul}

\begin{document}

   \title{Distance to the Globular Cluster M 3 from the Infrared Surface Brightness Technique applied to RR Lyrae stars}
   \titlerunning{The IRSB technique for RR Lyrae stars from M 3}
   \authorrunning{Zgirski, Gieren, Pietrzyński et al.}

   \author{Bartłomiej Zgirski
          \inst{1},
	  Wolfgang Gieren
	  \inst{1},
	  Grzegorz Pietrzyński
	  \inst{2},
	  Gergely Hajdu
	  \inst{2},
	  Piotr Wielgórski
          \inst{2},
	  Marek Górski
	  \inst{2},
	  Jesper Storm
	  \inst{3},
	  Nicolas Nardetto
          \inst{4},
	  Alexandre Gallenne,
	  \inst{5}
	  Garance Bras
	  \inst{6},
	  Pierre Kervella
	  \inst{6, 7},
	  Paulina Karczmarek
	  \inst{2},
	  Weronika Narloch
	  \inst{2}
          }

   \institute{Universidad de Concepción, Departamento de Astronomía, Casilla 160-C, Concepción, Chile\\
              \email{bzgirski@astro-udec.cl}
	     \and
             Nicolaus Copernicus Astronomical Center, Polish Academy of Sciences, Bartycka 18, 00-716 Warszawa, Poland
	     \and
	     Leibniz-Institut für Astrophysik Potsdam (AIP), An der Sternwarte 16, D-14482 Potsdam, Germany
	     \and
             Université Côte d’Azur, Observatoire de la Côte d’Azur, CNRS, Laboratoire Lagrange, France
	     \and
	     Instituto de Alta Investigaci\'on, Universidad de Tarapac\'a, Casilla 7D, Arica, Chile
	     \and
	     LESIA, Observatoire de Paris, Université PSL, CNRS, Sorbonne Université, Université Paris Cité, 5 place Jules Janssen, 92195
Meudon, France
             \and
	     Universidad de Chile, Departamento de Astronomía, Camino el Observatorio 1515, Santiago, Chile
	     }

  \abstract
   {The Infrared Surface Brightness (IRSB) technique is a specific application of the Baade-Wesselink method. Given a proper calibration, well covered optical and near-infrared photometry, as well as radial velocity curves, it allows for estimation of distances to individual pulsating stars and determination of their mean radii. Even though observationally demanding, it offers a way of distance determinations to systems of pulsating stars that is independent of period-luminosity relations. The technique is fully empirical and does not depend on stellar atmosphere models.}
   {The goal of the work is to test the precision of distance determinations to individual RR Lyrae stars and to their host system as a whole using the IRSB technique for a relatively distant globular cluster M 3 (NGC 5272). We also aim to determine mean radii and period-radius relations for these stars in order to compare them with the existing theoretical prediction and empirical estimations for the field stars from the solar neighborhood.}
   { We use data available in the literature and the calibration of the IRSB technique based on the RR Lyrae stars from the solar neighborhood we published previously in order to determine distances to 14 RR Lyrae stars in the globular cluster M 3. We apply the IRSB technique as previously presented in our work for RR Lyraes from the solar vicinity with an extra determination and implementation of phase shifts between observables due to the phase incoherence of data gathered at different epochs. We study the impact of the selection of the fitting procedure (bisector v.s. the least squares fit) on the results. We apply five different empirical surface brightness-color relations from the literature in the analysis.}
   {We obtained a mean distance to M 3 of $r_{M 3}=\left(10.07 \pm 0.19 \pm 0.29\right) \,kpc$ that corresponds to a distance modulus $\mu_{M3}=(15.015 \pm 0.041 \pm 0.063) \,mag$ and a 7\% scatter of individual stellar distances for 14 RR Lyrae stars in M 3. We received a very good agreement between the two fitting techniques, with the bisector proving to be a biased estimator while the least squares fit yielding slightly larger uncertainties. We also determined mean stellar radii for pulsators from the sample with a precision of $0.5\%$ and obtained excellent agreement with a theoretical prediction of the period-radius relation for RRab stars available in the literature.}
   {}

   \keywords{Stars: distances, Stars: oscillations, Stars: variables: RR Lyrae, Infrared: stars}

   \maketitle

\section{Introduction}
The Infrared Surface Brightness (IRSB) technique is a specific application of the Baade-Wesselink (BW) method (\citealt{Baade}, \citealt{Wesselink}) that allows to trace changes of the size of a radially pulsating star. Purposes of such analysis may vary, but usually they come down to the determination of distances to individual pulsating stars through the simple ratio of the stellar linear diameter and angular diameter. The method also allows for determination of mean stellar radii. While variations of the radius can be derived from the integration of the radial velocity (RV) curve obtained using a spectrograph, the angular diameter is determined for the corresponding pulsation phases simultaneously using an independent observing method. In principle, stellar angular diameters can be measured directly through interferometry but this is an observationally and technologically demanding type of observations. However, as originally noticed by \cite{BARNES}, empirical relations between photometric magnitudes and stellar angular diameter are possible to be calibrated. These are known in the literature as the surface brightness - color relations (SBCR). In such approach, interferometric observations are performed only for a calibrating sample of a given stellar species in order to calibrate a SBCR. Later, the relation is applied to estimate angular diameters of objects where interferometry is not available. Such reasoning was applied for the detached eclipsing binary systems of the Magellanic Clouds, obtaining the most precise distances currently available for these galaxies (1\% for the LMC and 2\% for the SMC, \citealt{LMC-DEB}, \citealt{SMC-DEB}).

Analyses performed in the 90s for classical Cepheids by \cite{WELCH} and \cite{FG} showed that, among different combinations of considered photometric passbands, the most precise estimation of angular diameters is obtained based on SBCRs that depend on the optical $V$ band and the $(V-K)$ color, where $K$ is the broad band centered on $2.16 \,\mu m$. The use of the near-infrared band allows for a better sensitivity to effective temperature while minimizing the influence of the gravity and line blanketing on the angular diameter determinations \citep{DIBENEDETTO}. SBCRs as functions of $V$, $(V-K)$ have been proven to be practically independent on metallicity and they depend very little on the reddening (\citealt{THOMPSON}, \citealt{STORM2004}, \citealt{LMC-DEB}, \citealt{IRSB-GAL}\footnote{that work devoted to the calibration of the IRSB technique for RR Lyrae stars will be cited here under the acronym 'Z24'.}). 

The IRSB technique for pulsating stars is derived from all these considerations and it has been applied in the analysis of classical Cepheids, e.g., \cite{GFG97}, \cite{BW-MET}, where the authors determined distances to individual pulsators and calibrated period-luminosity (PL) relations for them. More recently, we used the IRSB technique in the analysis of BL Her stars \citep{IRSB-T2CEP} and RR Lyrae stars (Z24) in the solar neighborhood. It is a fully empirical approach to the BW method that relies on SBCR calibrations resulting from observations and it is independent of the stellar atmosphere models.

The parameter known as the $p-$ factor that is a ratio of the actual alteration of the stellar radii to the integral of the observed RV between given pulsation phases is crucial in every calibration and application of the BW method. As described by \cite{NPFAC}, $p$ may be divided into different factors of a product, including the fact that the measured RV of a given part of the stellar surface is a projection of the pulsational velocity of the surface projected on the line of sight. Another components of the $p-$ factor correspond to the ratio of gas velocity of the photosphere and of the line forming region and the ratio of the pulsational and the gas velocity of the photosphere.

It is important to note that calibrations based on different applications of the BW method yield different values of the $p-$ factor \citep{IAUS}. In order to avoid the commitment of a systematic error in the distance estimation, it is thus pivotal to apply the method in the same exact way as it was done for its calibration. We can clearly see this comparing the analyses performed by us (\citealt{GARANCE}, Z24) for RR Lyrae stars from the solar neighborhood using the SPIPS \citep{SPIPS} code and the IRSB method. While we obtained similar scatter of $p-$ factors ($6\%-7\%$) using both techniques, the mean values of $p$ are systematically shifted between each other (with the mean value of $p=1.24$ from SPIPS and $p=1.39 - 1.45$ from IRSB, depending on a SBCR). It highlights that various applications of the method are sensitive to specific components of the $p-$ factor in a different way.

According to a long-established view \citep{PULS}, RR Lyrae stars trace the old stellar population and allow determining distances to systems where young stellar distance indicators, like classical Cepheids, are usually not present, such as dwarf spheroidal galaxies and globular clusters. Even though their picture of being exclusively members of the metal-poor Population II has been questioned in the literature (\citealt{AB}, \citealt{ZHANG}), there is no empirical proof for the hypothesis that a large part of the population of nearby RR Lyrae stars is made out of members of Population I with metallicities similar to solar (e.g. determinations for nearby RR Lyrae stars of \citealt{CRESTANI} prove their low metallicity). Moreover, simulations of the evolution of metal-rich binaries performed by \cite{BEP-2017} showed that the contamination of regular RR Lyrae stars by metal-poor and low-massive Binary Evolution Pulsators \citep{BEP}, that imitate the RR Lyrae variability, is at the level of less than $1\%$.
While distances based on RR Lyraes are determined using period-luminosity(-metallicity) relations typically (e.g. \citealt{KBAND}, \citealt{RRLPLZ}, \citealt{WER}), the BW analysis provides independent, though more observationally demanding, mean of distance estimations for these stars. We made a broader review of the literature related to the BW method for RR Lyrae stars in the introduction to our calibration paper (Z24). In the context of this paper, it is important to mention the work of of \cite{JURCSIK0} that dealt with a task of BW distances to 26 RR Lyrae stars from M 3 based on the optical $(V-I)$ color and the stellar atmosphere models of \cite{CastelliKurucz}. In that work, a theoretical uniform value of $p=1.35$ \citep{NARDETTO2004} computed for a model of the prototype classical Cepheid, $\delta$ Cep, was assumed. The authors reported a splendid scatter of individual stellar distances of $3\%$. 

The goal of this work is to determine distances and mean radii of RR Lyrae stars from the globular cluster M 3 (NGC 5272) using the previously calibrated IRSB technique and data available in the literature in order to verify the precision of the technique applied to a spatially confined system and investigate the period-radius (PR) relation for RR Lyraes from the cluster.

\section{Data}

We took optical light curves in the $V$ band and RV curves from the previously mentioned work of \cite{JURCSIK0} and the $K_s$ band photometry of RR Lyrae stars in M 3 published by \cite{KBAND} that is calibrated in the 2MASS system \citep{2MASS}. In order to minimize the impact of decoherence of periodic data having periods determined with finite precision, we used temporally-compact span of epochs for each observable\footnote{By using the term 'observable', we mean $V$, $K$ magnitudes and RV measurements.}. While in the case of $V$ and $K$ magnitudes, we used all available data, radial velocities for M 3 RR Lyraes were obtained in two observing campaigns in 2000 and in 2012. These observations were performed using two different spectrographs. Observations from 2000 were gathered by Hydra installed on 3.5 m WYIN telescope at the Kitt Peak National Observatory (\citealt{HYDRA}, \citealt{HYDRA_RRL}), while in 2012, Hectoechelle installed on 6.5 m MMT in Arizona \citep{Hectoechelle} was used. Thus, assuming a typical pulsation period of an RR Lyrae from our sample of $P=0.6$ d, the two observing campaigns are separated by more than 2600 pulsation cycles, while each observing campaign was completed in less than 80 typical pulsation cycles. We decided to use data just from one spectrograph, preferring more precise RV mesurements from Hectoechelle. However, in the case V 51, V 81, and V 84, these did not have good enough coverage to construct a proper empirical model of the RV curve. In these cases, we only used data from Hydra.

We cross-matched data from \cite{JURCSIK0} and \cite{KBAND} works and obtained datasets available for analysis using the IRSB technique for 14 RRab stars.

\section{Calibration of the technique}

In our previous work (Z24), we delivered empirical calibrations of the IRSB technique based on 8 RRab stars in the solar vicinity having their accurate\footnote{As shown in \cite{RRLPLZ}, near-infrared absolute magnitudes of nearby ($r<1.5\,kpc$) RR Lyrae stars obtained based on Gaia DR3 parallaxes are coherent with the photometry of these stars located in the LMC and the accurate distance to that galaxy based on detached eclipsing binaries \citep{LMC-DEB}. However, we observed a zero point offset between our PL relations and those found in the literature that are based on more distant stars and the same source of parallaxes.} distances determined by the Gaia mission \citep{GAIAEDR3} for four different SBCRs. While there is no SBCR calibrated for RR Lyrae stars available in the literature, we compared $p-$ factor values and mean radii resulting from the application of three SBCRs for dwarfs and subgiants (\citealt{KER_SBCRI}, \citealt{GRACZ-SBCR}, \citealt{SALSI}, calibrated mean $p-$ factor values of 1.44, 1.39, 1.45, respectively) and one SBCR for classical Cepheids (\citealt{KER_SBCRII}, calibrated mean $p=1.40$). We also found relations between $p$ and the pulsation period ($P$) of RRab stars from our sample. Even though our calibrating sample is not large and the obtained uncertainties of relations between $\log P$ and $p$ are considerable, our technique-specific calibration is fully empirical, based on very good-quality data and the state-of-the-art accurate Gaia parallaxes of RR Lyraes being closer than 1.5 kpc from the Sun. The calibration gives advantage over $p$ values previously assumed in many applications of the BW method, including for RR Lyrae stars.

In the original calibration paper (Z24) we provided explicitly parameters of only two, extreme $p(P)$ relations. Table \ref{tab:pP} provides parameters of these relations and derived mean $p-$ factors for all considered SBCRs and the two considered types of linear fit.\footnote{In the original calibration work (Z24), we only provided the calibration based on the bisector fit. We report calibration resulting from the LS fit and based on the same sample as in the Z24 for the first time here.}. In Table \ref{tab:pP}, we also give an additional calibration based on the recent SBCR of \cite{SBCR-B2025} for classical Cepheids, which is not included in the Z24 paper. In the original work, we only gave parameters of linear relations between the angular diameter and the integral of radial velocities based on the bisector fit. We additionally provide relations' parameters based on the least squares fit in the same Table \ref{tab:pP}.

\section{Application of the technique}

We provided a detailed description of the technique we apply in the Z24 work. In order to phase the data, we adopt pulsation periods of M 3 RR Lyraes from the \cite{PERIOD} paper. We fit Akima splines \citep{akima} to the $V$ light curve and the RV curve. Then, we calculate $(V-K)$ color\footnote{The SBCR of \cite{SALSI} and \cite{GRACZ-SBCR} are calibrated for the 2MASS $K_s$ band already. However, when applying the SBCRs of \citeauthor{KER_SBCRI} (\citeyear{KER_SBCRI}, \citeyear{KER_SBCRII}), we had to transform the $K_s$ band photometry onto SAAO $K$ magnitudes using transformation equations given in the work of \cite{KOEN}, just like we did it in the Z24 work. In this sense, the $(V-K)$ correponds here to the color that an applied SBCR is based on, with the 2MASS $K_s$ being one of the possible calibrations of the $2\,\mu m$ $K$ magnitude.} for epochs corresponding to the measured $K$ band magnitudes. Having that, we perform SBCR-based estimations of angular diameters for these epochs where $V$ band magnitude is interpolated using the spline model. $K$ band light curve is always treated as a collection of discrete measurements, except when used to estimate the scatter of $K$ band magnitudes. For this task, we fit amplitudes and zero points of the Fourier series-based templates published by \cite{template}, and calculate the scatter of measurements around the fit.

Essentially, our approach comes down to the linear fit of the following relation:

\begin{equation}
\theta(x)=p \varpi x + \theta_0
\label{eq:angdia}
\end{equation}

where $\theta$ is the angular diameter, $x(\phi)=-2 \int_{0}^{\phi} [v_r(\phi') -v_{r,0}] d \phi'$ is the integral of RV curve ($v_r$ being the measured RV, $v_{r, 0}$ being the mean, systemic RV) over the pulsation phase $\phi$. The fitted slope is a product of the $p-$ factor and the stellar parallax ($\varpi$), while the intercept corresponds to the angular diameter for $\phi=0$ defined as the phase of maximum brightness in the $V$ band. In our analysis, we consider two types of linear fits to the above relation - the bisector method \citep{BISECTOR} and the ordinary least-squares fit (OLS, LS). In this paper, we report results based on these two types of estimation.

While in the Z24 paper we assumed $\varpi$ from Gaia in order to calibrate $p$, in the current work we apply SBCR-specific $\log P - p$ (Pp) relations in order to derive distance $r=1/\varpi$.

Additionally, we determine mean radii of the studied RR Lyraes from M 3:
\begin{equation}
    \langle R\rangle=\frac{\langle r \rangle \theta_0}{2}+ \langle \Delta R \rangle = \frac{\langle r \rangle \theta_0}{2} - \langle r \rangle p \varpi \int\displaylimits_0^1{\int\displaylimits_{0}^{\phi}{\left[v_r(\phi')-v_{r,0}\right]d \phi'} d \phi}
    \label{eq:radius}
\end{equation}

where $\langle r \rangle$ is the mean distance to stars from our sample that is generally SBCR-dependent. The above equation follows directly from the Equation 6 of the Z24 work. Here, the mean stellar radii of individual stars depend on the cluster distance, previously estimated as a mean of individual stellar distances, $\langle r \rangle=1/\varpi$, and the linear fit slope $p \varpi$ is written explicitly. Such approach allows us to obtain better precision of the stellar radii determinations.  Derivation of mean radii accordingly to Equation \ref{eq:radius} relies on parameters of the $\theta (x)$ relation (Equation \ref{eq:angdia}). In this formulation, it does not depend directly on the $p-$ factor.

 Pulsation periods adopted from \cite{PERIOD} provide good phasing of the data. However, the fact the periods are short, never determined with infinite precision, and drifting slowly with time, what is characteristic for RR Lyrae variables, causes decoherence between batches of datasets that were taken at time-separated epochs, differing by many pulsation periods between each other. In order to account for that decoherence, we independently estimate $\Delta \phi_{V-K}$ and $\Delta \phi_{RV-K}$ values that correspond to the phase shift between $V$, $K$ light curves and between RV curve and $K$ light curve, respectively. Namely, we probe the space of forementioned phase shifts and corresponding scatter around the bisector fit to the relation from the Equation \ref{eq:angdia} with the two independent phase steps of 0.002. Phase shifts that yield the smallest scatter of points around the fitted relation between the angular diameter and the integral of radial velocities, provide the adopted $\Delta \phi_{V-K}$ and $\Delta \phi_{RV-K}$ values. The adoption of pulsation period change rates published by Jurcsik+(2012) does not yield proper coherence of the empirical curves in some cases. When using period change rates determined for stars from our sample, we generally obtained phase shifts that resulted in a wrong correspondence between angular diameters and integrals of RV curves, as well as between different empirical curves ($V$, $K$, RV). In a more general context of the used technique, the development of the independent phase shifts' determination is more robust as it allows to apply the method for stars that do not have a long history of observations that could allow for the analysis of their period changes. Our approach also follows the reasoning present in the previous works devoted to the IRSB technique (such as \citealt{GFG97}), where authors determined phase shifts based on the minimization of the scatter around the relation between the stellar angular diameter and the integral of the RV curve.

\section{Analysis and results}

\begin{figure}
\includegraphics[width=0.5\textwidth]{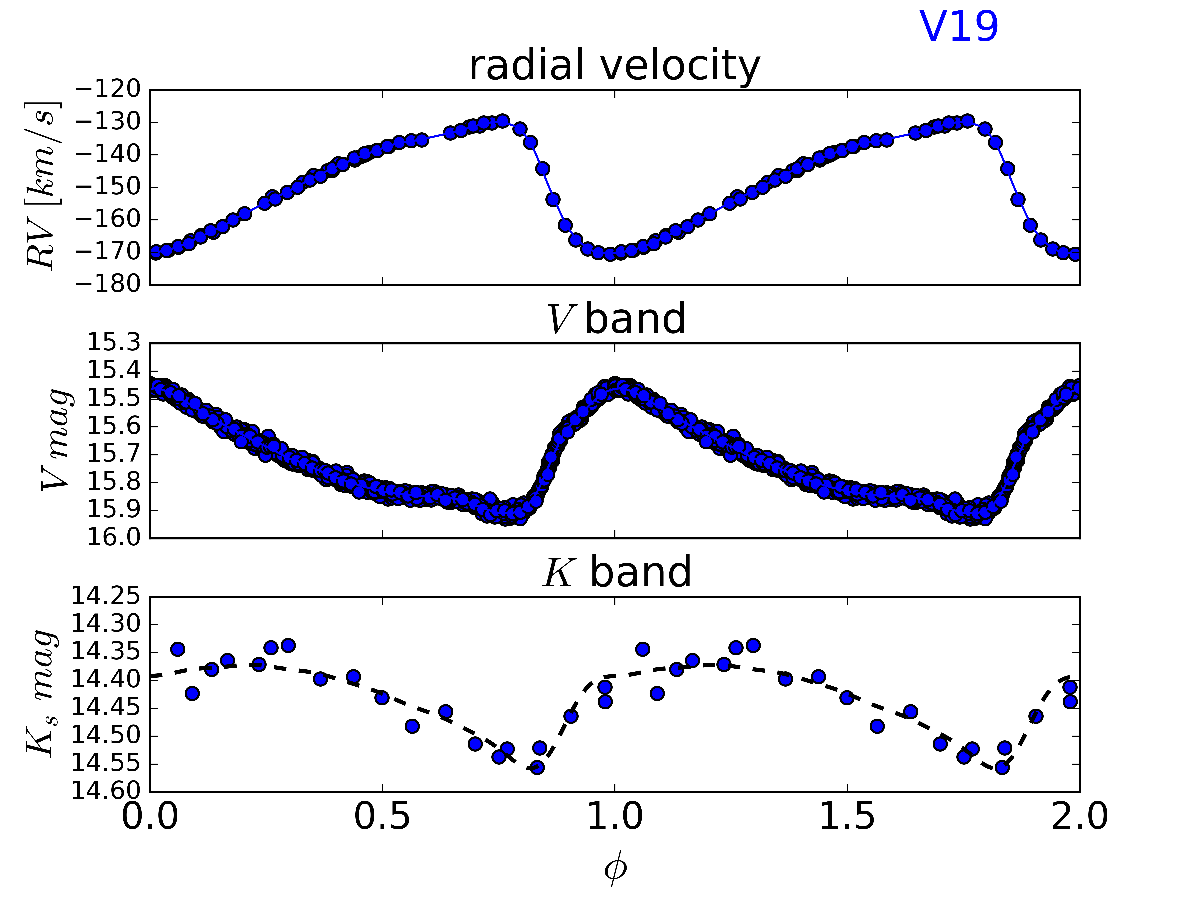}
\caption{Light and RV curves for V 19 phased with zero points of the phase determined in this work. Plots corresponding to all stars from our sample are available on Zenodo.}
\label{CURVE}
\end{figure}

In the paper, we present plots related to the analysis for the variable V 19 based on the SBCR of \cite{KER_SBCRI}. The collection of plots corresponding to all stars from our sample and all five applied SBCRs is available on Zenodo. All plots present data and results where phase shifts between observables have been applied already. Figure \ref{CURVE} presents light and RV curves for V 19 together with the Akima splines modeling the variation of RV and $V$ band magnitude (model for the $V$ magnitude is not visible due to the dense coverage of the light curve). The bottom panel of the plot depicts the $K$ band light curve with the template fit represented by a dashed curve.

\begin{figure}
\includegraphics[width=0.5\textwidth]{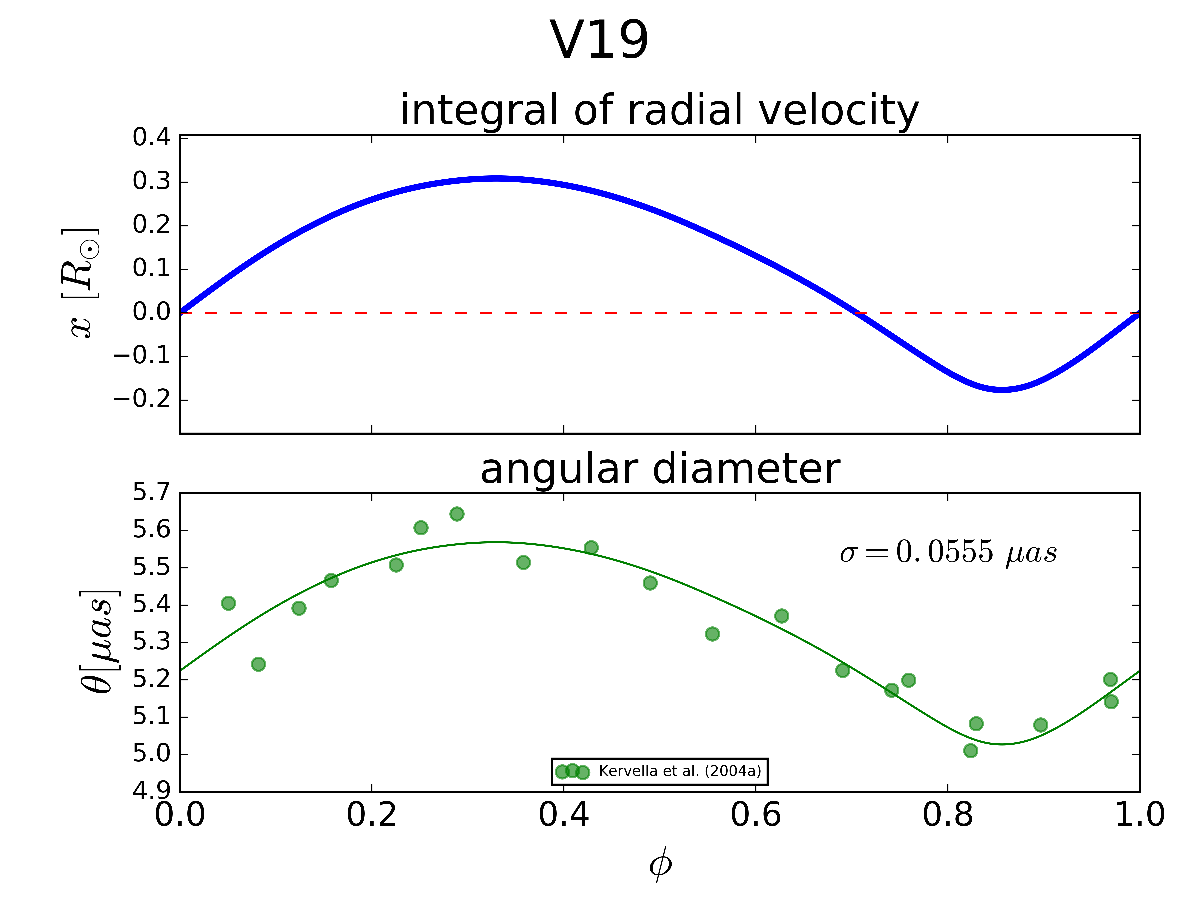}
\caption{Top: Integral of the RV curve. Bottom: The course of the angular diameter based on one of the five used SBCRs for V 19. 
Curve in the bottom panel is not a fit, but the curve from the top scaled by $2 p \varpi$ - the slope resulting from the bisector fit. Plots corresponding to all stars and all assumed SBCRs are available on Zenodo.}
\label{RADIA}
\end{figure}

\begin{figure}
\includegraphics[width=0.5\textwidth]{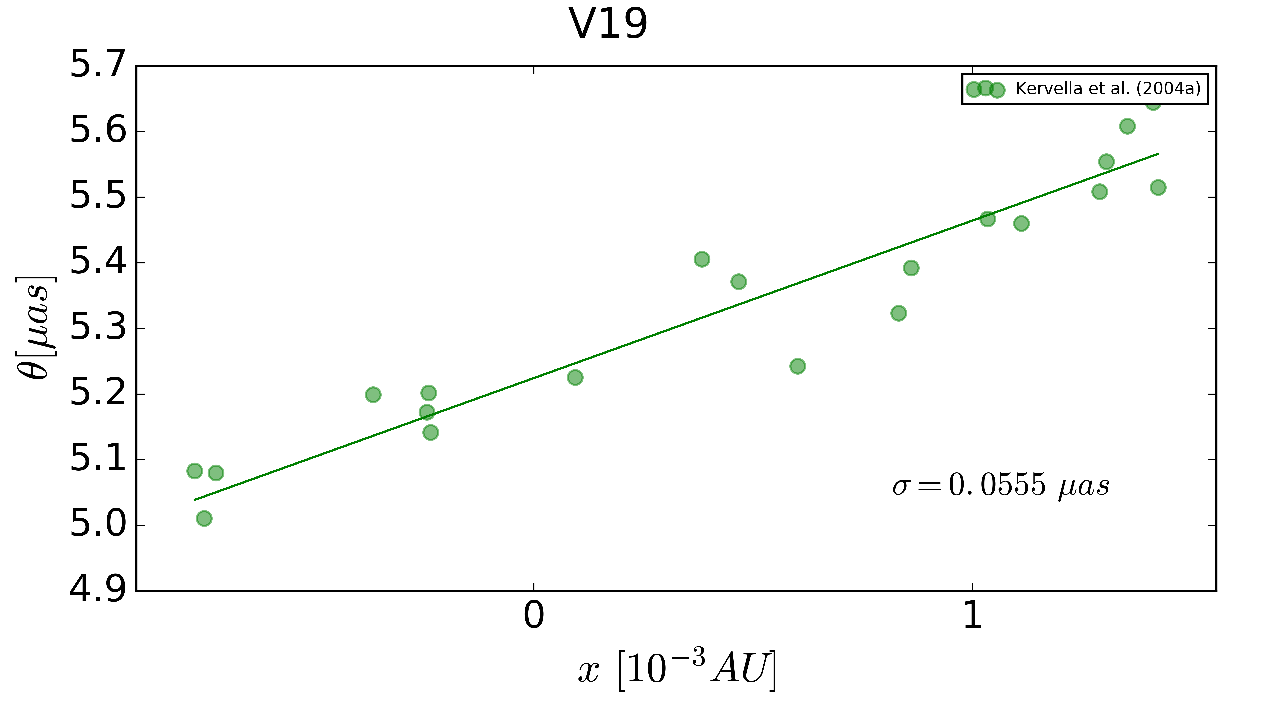}
\caption{Linear bisector fit of a relation between the integral of RV and the angular diameter for V 19. Plots corresponding to all stars and all assumed SBCRs are available on Zenodo.}
\label{BISEC}
\end{figure}

Table \ref{tab:products} presents estimated values of $\Delta \phi_{V-K}$ and $\Delta \phi_{RV-K}$ phase shifts together with the $p \varpi$ products and the corresponding fit uncertainties for all studied objects, depending on the used SBCR. We might see that formal fit uncertainties are significantly larger for the LS fit. Figure \ref{RADIA} presents the course of the integral of the RV curve (the upper panel) and the course of the angular diameter curve together with $\theta$ estimations for specific phases (lower panel). Figure \ref{BISEC} depicts the bisector fit between the two variables. The slope of the line depends the most on the extreme values of $\theta$ and $x$. Thus, it is important to have well-probed minima and maxima in Figure \ref{RADIA}, i.e., estimate accurately amplitudes of size variations\footnote{The ratio of amplitudes of the two curves in Figure \ref{RADIA} equals to $2p \varpi$. The curve in the bottom panel of that Figure actually corresponds to the curve from the upper panel scaled by the slope of the fitted relation from Figure \ref{BISEC} - the same, as in the calibration paper (Z24).}. Otherwise, we must rely on the extrapolation of the position of an extremum. In the case of three stars - V 55, V 81, and V 83, we are dealing with large gaps in their coverage of the $K$ band light curves. This results in the lack of the estimation of $\theta$ near the maximum size. We still decided to keep these objects in our sample as they have parts of their $K$ band light curves, including those corresponding to the minimal stellar size, relatively well-covered.

We performed distance determinations for individual stars from our sample through the calculation of the $\frac{p(P)}{p \varpi}$ ratio where the numerator is the value of the $p-$ factor for a given stellar pulsation period resulting from the calibration work (Z24) and the denominator is the slope of the fitted $\theta (x)$ relation. Table \ref{tab:dist} summarizes determinations of distances for all stars from the sample together with mean distances and the corresponding standard errors, depending on a SBCR and a fitting method.

We propagated uncertainties of derived distances and mean stellar radii using Monte Carlo simulations where we included the following error components: scatter of radial velocities and $V$ magnitudes around interpolating Akima splines, scatter of $K$ magnitudes around fitted second order Fourier fit, uncertainty of $p-$ factor for a given pulsation period (Table \ref{tab:pP}). The systematic uncertainty components include $V$ and $K$ magnitude uncertainty of photometric zero points ($0.01$ mag), and the formal systematic uncertainty of $E(B-V)$ reddening assumed as $0.01$ mag - even though the reddening for M 3 is extremely small, $E(B-V)=0.014$ mag (\citealt{JURCSIK0}, \citealt{HARRIS}). The impact of reddening on the slope of the $\theta (x)$ relation is also known to be very small compared to other factors of the total uncertainty (Z24). In the case of mean radii uncertainties, the variation of previously determined mean cluster distance was performed instead of the variation of $p-$ factor value in order to estimate one of the components of the systematic uncertainty.

The largely dominating component of distance uncertainties results from the scatter of the $K$ band photometry that in the case of the used data corresponds to around $0.02$ mag. In the case of V 19, the SBCR of \cite{GRACZ-SBCR} and the bisector fit yield the total simulated distance distribution rms of 870 pc. The scatter of $K$ magnitude measurements propagates into 830 pc of the total statistical distance error. The component associated with the uncertainty of the $p-$ factor is significantly smaller (250 pc), while the other components are negligible.

\begin{figure}
\includegraphics[width=0.5\textwidth]{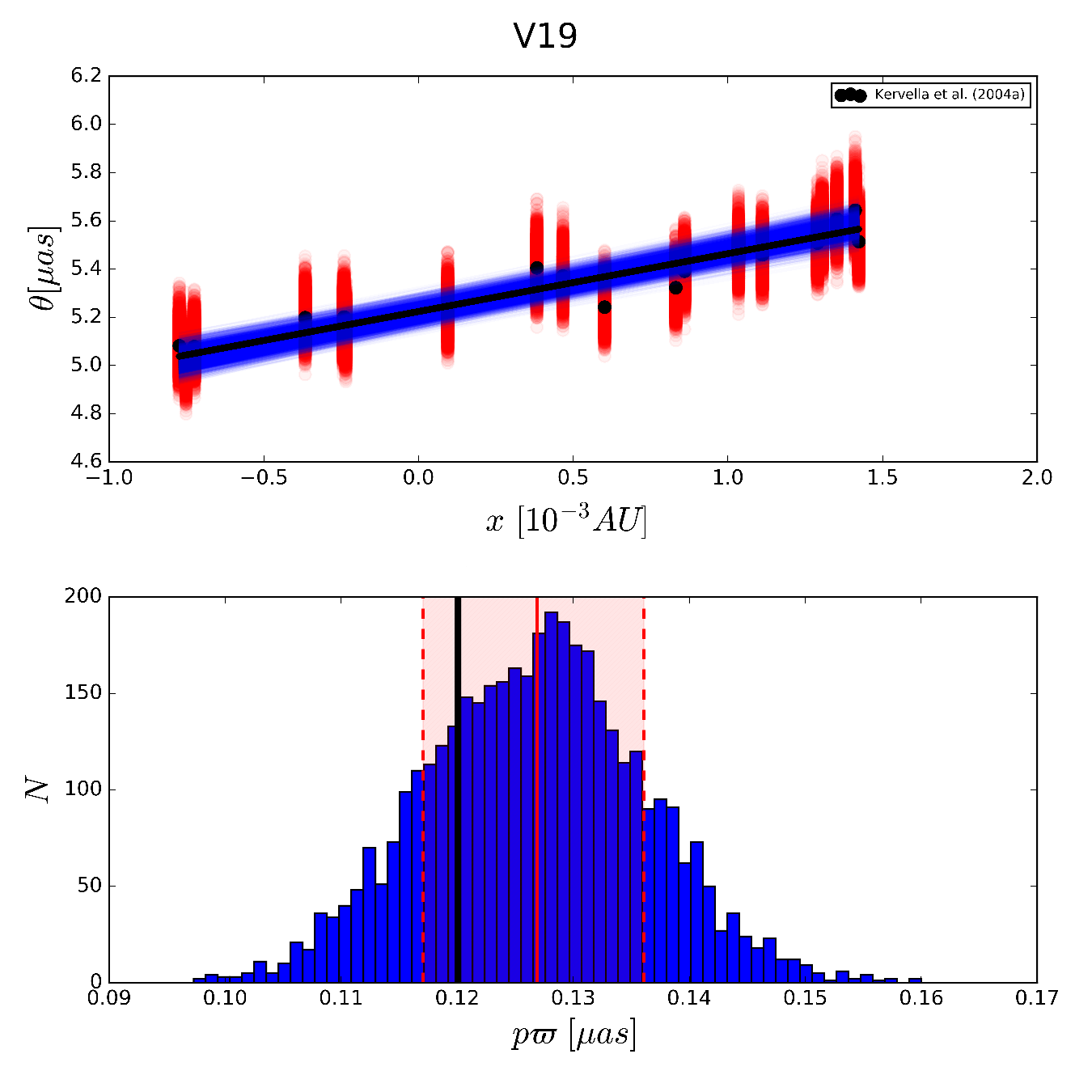}
\caption{Top: Bisector fits to MC-simulated data. Bottom: Corresponding distribution of fitted slopes, $p \varpi$ products, for V 19 together with marked $1 \sigma$ confidence interval. Distributions corresponding to all stars and all assumed SBCRs are available on Zenodo.}
\label{PW_BISEC}
\end{figure}

\begin{figure}
\includegraphics[width=0.5\textwidth]{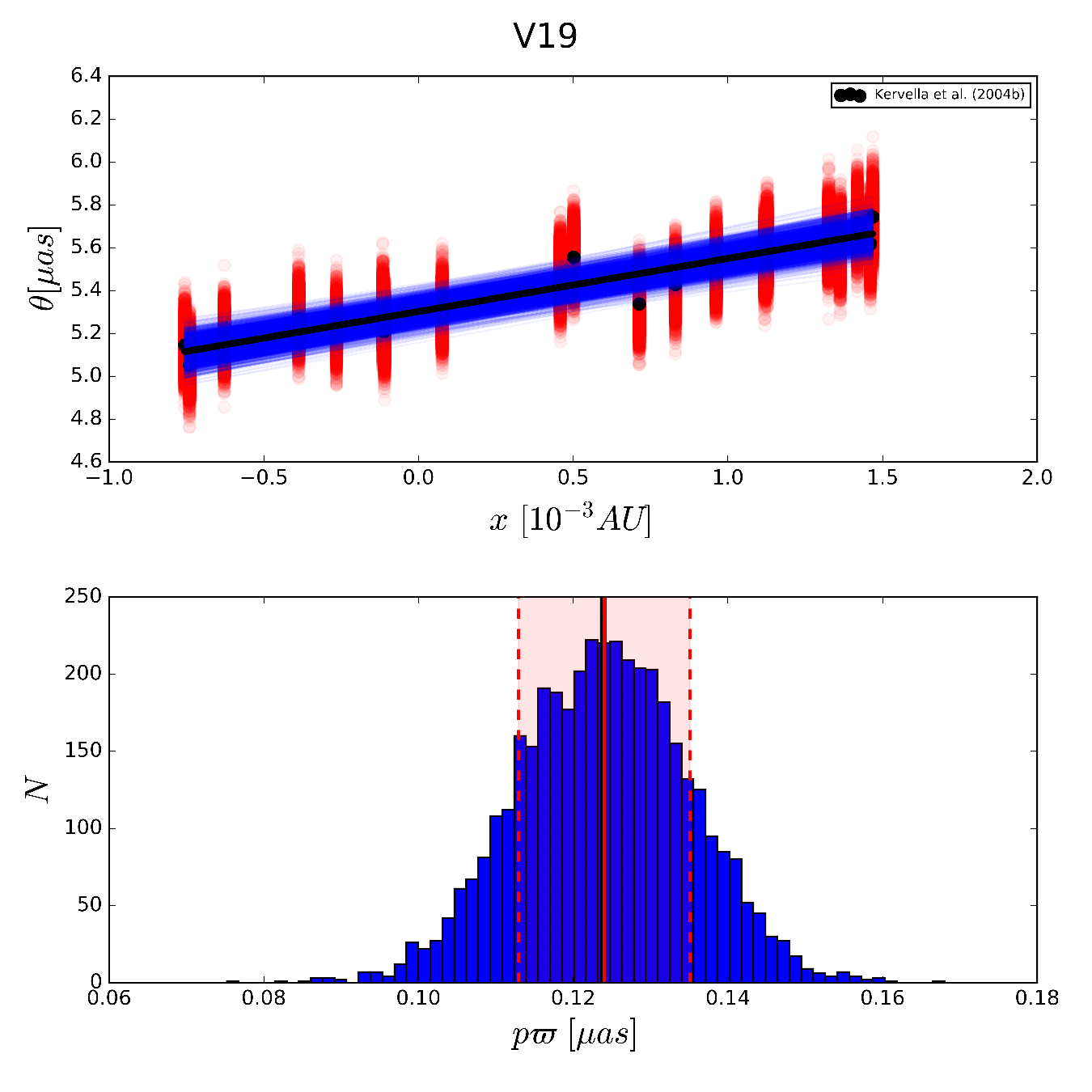}
\caption{Top: Least-squares fits to MC-simulated data. Bottom:Corresponding distribution of fitted slopes, $p \varpi$ products, for V 19 together with marked $1 \sigma$ confidence interval. Distributions corresponding to all stars and all assumed SBCRs are available on Zenodo.}
\label{PW_LS}
\end{figure}

In our simulated distributions of $p \varpi$ based on the bisector fit, we observe a bias between the directly calculated slope and the median (mean, mode) of the simulated distribution. We do not observe such bias when using the LS fit. Figures \ref{PW_BISEC} and \ref{PW_LS} depict fits and distributions of $p \varpi$ resulting from Monte Carlo (MC) simulations with black points and lines corresponding to the actual determination. The black vertical line in the lower panel corresponds to the measurement based on the original, unvaried data. The bisector bias is not that apparent in the calibration work (Z24) as the Gaia parallax uncertainty yields an error component of the derived $p-$ factor that is similar to the component corresponding to the $K$ band magnitude scatter. 

However, the LS fit yields results that are less stable than those obtained based on the bisector. The spread of simulated distributions is larger, resulting in larger statistical uncertainties. We obtain the scatter of individual stellar distances of 7\% for the bisector and 7.5\% for the LS fit. Mean rms of simulated distributions corresponding to individual distance measurements is at the level of 6.5\% and 8\% of the distance for the bisector and for the LS estimator, respectively. 

Since the original calibration work (Z24) was based on the bisector fit and due to the better stability of solution, we still rely on the slope of the bisector fit as the basis of the preferred estimator of distance. Namely, when calculating the M 3 distance, we rely on the average of the five mean bisector-based distances for the all used SBCRs with the mean of the corresponding standard errors being the statistical uncertainty of our estimation ($0.19\,kpc$). The shift between mean cluster distances based on the bisector and the unbiased LS fit ($\approx 50 \,pc$, i.e., 0.5\%) serves us as the estimation of a component of the systematic uncertainty of the distance determination to M 3. As we may see in Table \ref{tab:dist}, the maximum observed distance shift resulting from the choice of SBCR is of the order of $\approx 110 \,pc$, i.e., 1.1\% in the case of the bisector. The component of the systematic distance uncertainty associated with photometric zero points is of the order of $0.9\%$ and the component associated with the uncertainty of the $E(B-V)$ reddening is of the order of $0.4\%$ of the estimated cluster distance. In total, these factors constitute the mean total systematic distance uncertainty of $2.9\%$.
Finally, our estimation of the M 3 distance is: 
\begin{equation}
r_{M 3}=\left(10.07 \pm 0.19 \pm 0.29\right) \,kpc
\end{equation}

which corresponds to the true distance modulus of $\mu_{0, M3}=(15.015 \pm 0.041 \pm 0.063) \,mag$. The statistical uncertainty of $0.19\,kpc$ corresponds to the standard deviation of individual distances of $0.71\,kpc$ for our sample of 14 stars.

\begin{figure*}
\includegraphics[width=1.0\textwidth]{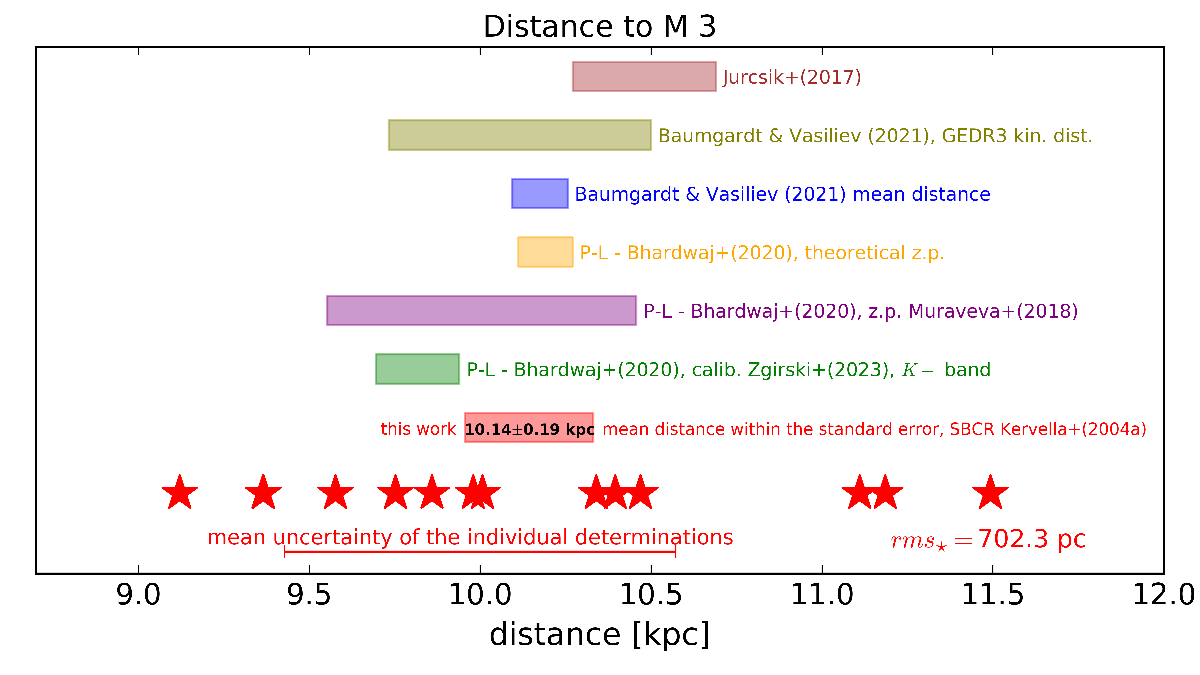}
\caption{Comparison of distances to M 3 within their statistical uncertainties from the literature with the result from this work for the SBCR of \cite{KER_SBCRI} and the bisector fit. Plots corresponding to all assumed SBCRs are available on Zenodo.}
\label{DISTS}
\end{figure*}

Figure \ref{DISTS} depicts the M 3 distance together with individual stellar distances resulting from the SBCR of \cite{KER_SBCRI}. They are compared with some of the recent M 3 distances available in the literature. The forementioned work of \cite{JURCSIK0}, which is also the source of optical photometry and radial velocities used in this work, yielded $r_{M 3}=10.48 \pm 0.21 \,kpc$. In their compilation of Galactic cluster distances, \cite{BV} report, among others, the Gaia EDR3 kinematic distance to M 3, $r_{M 3}=10.116 \pm 0.384 \,kpc$ and a distance that is a mean of 38 literature determinations, $r_{M 3}=10.175^{+0.082}_{-0.081} \,kpc$. The paper of \cite{KBAND} that is also a source of the near-infrared photometry utilized in this work, presents distance to M 3 based on PL relations for RR Lyrae stars. Originally, the authors reported the true distance modulus of $\mu_{0, M3}=(15.041 \pm 0.017 \pm 0.036) \,mag$ based on the theoretical zero points for $JHK_s$ bands from the work of \cite{MARCONI2015}. They additionally present distance estimation based on the empirical zero points of \cite{Muraveva} based on Gaia DR2 parallaxes, $\mu_{0, M3}=(15.001 \pm 0.098 \pm 0.121) \,mag$. We used coefficients of PL relations reported in \cite{KBAND} and our sample of nearby Galactic RR Lyrae stars from \cite{RRLPLZ}, which had their absolute magnitudes determined based on Gaia DR3 parallaxes. We obtained M 3 true distance moduli of (statistical errors) $\mu_{0, M3}^J=(14.973 \pm 0.034) \,mag$ and $\mu_{0, M3}^K=(14.955 \pm 0.027) \,mag$ for the $J$ and the $K$ band, respectively. All of the above results are in agreement with the distance determination to M 3 in the current work.

\begin{figure*}
\includegraphics[width=1.0\textwidth]{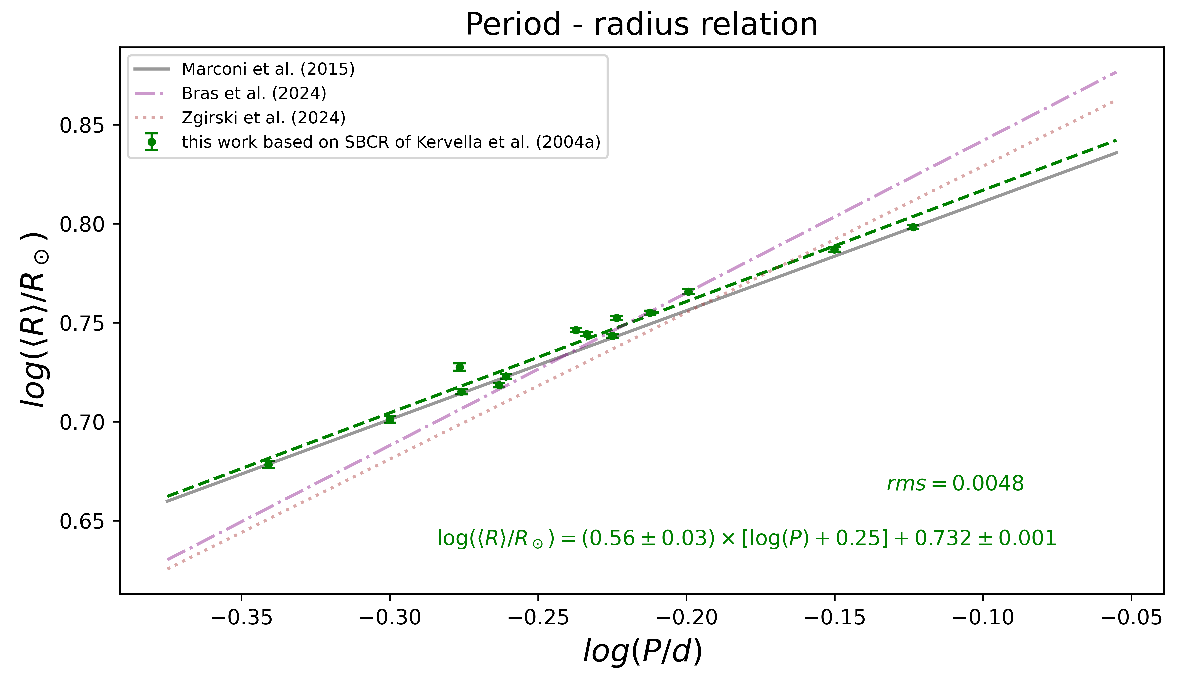}
\caption{Period-radius relation derived in this work based on the SBCR of \cite{KER_SBCRI} and the bisector fit compared with the theoretical relation of \cite{MARCONI2015} and the two relations derived for RR Lyrae stars from the solar neighborhood. Plots corresponding to all assumed SBCRs are available on Zenodo.}
\label{PR}
\end{figure*}

Based on Equation \ref{eq:radius}, we also determine mean radii of M 3 RR Lyraes from our sample. Table \ref{tab:radii} contains results of these determinations for all considered SBCRs and the two estimation types. Considering the bisector estimator of $\theta (x)$, Table \ref{tab:PR} presents coefficients of the fitted PR relations and Figure \ref{PR} depicts the ($\log P$ - $\log R$) plot for the SBCR of \cite{KER_SBCRI}, a fitted linear relation, and three literature PR relations of \cite{GARANCE}, Z24, and \cite{MARCONI2015}. 

For all considered SBCRs we obtain a remarkably low scatter around the fitted PR relation that corresponds to the relative precision of the radius estimation of around 0.5\%. The relevant components of the systematic uncertainty of the determined mean stellar radii are associated with the uncertainty of the parallax (distance), the photometric zero points, and the $E(B-V)$ reddening. Propagating $0.01\,mag$ $V$, $K$ zero point and $E(B-V)$ errors, we obtain the corresponding components of the systematic uncertainty of mean stellar radii of $0.01 R_{\odot}$, $0.035 R_{\odot}$, and $0.016 R_{\odot}$, respectively. The $0.2\,kpc$ error of the cluster distance corresponds to the systematic mean radius uncertainty of $0.11 R_{\odot}$. In total, we estimate the total systematic uncertainty of the mean stellar radii of $0.17 R_{\odot}$, i.e., around $3\%$ of the typical radius of RR Lyrae star.

We note a splendid agreement of our PR relations with the theoretical \cite{MARCONI2015} PR relation, much better than for PR relations for RR Lyrae stars from the solar neighborhood of \cite{GARANCE} and Z24 that were based on Gaia DR3 parallaxes. \cite{MARCONI2015} reports the following relation for RRab stars: $\log \left({\left<R\right>}/R_{\odot}\right)=(0.55 \pm 0.02) \log P +0.866 \pm 0.003$. Taking the pivot $logP_0=-0.25$, like in our relation, the intercept of that relation corresponds to $b=0.729$. That relation is in especially good agreement with our PR relations resulting from the SBCR of \cite{SBCR-B2025} for classical Cepheids, $\log \left({\left<R\right>}/R_{\odot}\right)=(0.552 \pm 0.026) \left(\log P +0.25 \right) +0.7268 \pm 0.0014$ and $\log \left({\left<R\right>}/R_{\odot}\right)=(0.555 \pm 0.027) \left(\log P +0.25 \right) +0.7288 \pm 0.0014$ for the bisector and the least squares fit of parameters of the $\theta(x)$ relation, respectively.

\section{Discussion and conclusions}

In this work, we obtained individual distances and mean radii for 14 RR Lyraes in M 3 as well as the estimation of the mean distance to the cluster based on the data available in the literature. Our determinations are based on a calibration dedicated to the same exact, fully empirical, application of the method, which is crucial given the fact that any $p-$ factor calibration is also application-dependent and distance scales as $r \propto p$ given a $p \varpi$ solution. We report a scatter of individual stellar distances of around $7\%$. According to our Monte Carlo simulations of the error propagation, the component of the statistical uncertainty of an individual distance determination that is related to the scatter of the $K$ band photometry is overwhelmingly large compared to other sources of the total statistical uncertainty. The size of the M 3 system itself is of the order of $100\,pc$ \citep{M3size} that corresponds to $1\%$ of the cluster's distance.
We obtained a statistical uncertainty of the cluster distance calculated as a standard error of the mean of individual distances of around 2\%. We applied the same five SBCRs in the calibration and in the determinations performed for the purpose of this work, with the recent SBCR calibration \citep{SBCR-B2025} for classical Cepheids based on the optical photometry included for the first time in our IRSB analysis.

We found that the slope of bisector line corresponding to the estimation of the $p \varpi$ product is a biased estimator through the Monte Carlo simulations yielding possible values of derived individual stellar distances. This is possibly associated with the uncertainties of the angular diameter estimation being significantly larger than uncertainties of the integral of the RV curve.
The slope resulting from LS fitting of a line is not biased, but more unstable, yielding larger scatter and statistical uncertainty than the one resulting from the bisector fit. We estimated the systematic distance uncertainty of $290 \,pc$ - $2.9\%$ through the comparison of the estimation based on the bisector fit with the result yielding from LS ($0.5\%$ - no significant offset between distances resulting from the two types of estimators is observed), the choice of SBCR ($1.1\%$ in the case of the preferred bisector fit), the photometric zero points uncertainty (up to $0.9\%$), and the uncertainty of the $E(B-V)$ reddening ($0.4\%$). Deriving the LS distance based on the bisector calibration of the method, we would obtain a systematic mean distance increase relative to the bisector distance of $5\%$ instead of $0.5\%$. It again shows that any determination should closely follow the same recipe as the calibration of the method. Even though the calibration based on the LS provided in Table \ref{tab:pP} is of lower quality than the calibration based on the bisector, it is the proper calibration for any determination of distance based on the LS version of the technique.

We also performed an exercise of the distance determination assuming constant value of $p$ for a given SBCR and fitting method. The average $p$ values based on the calibration work are available in the last column of Table \ref{tab:pP}. Summary of this analysis is depicted in Table \ref{tab:meandist}. The uncertainty of the mean distance with the assumption of the uniform $p$ value for a given SBCR and a fitting method is almost the same as the one resulting from the assumption of a linear Pp relation. Mean distances based on the $p$ independent of the pulsation period are on average slightly smaller than values resulting from Pp relations by $0.5\%$ and $0.2\%$ for the bisector and the LS versions of the technique, respectively.

We generally obtained good agreement within statistical uncertainties with other distance determinations to M 3 that are available in the literature. Regarding the determination based on the semi-empirical application of the Baade-Wesselink method \citep{JURCSIK0}, the authors obtained a value that is in agreement with the distance obtained in this work, but with the expected value of the mean distance that is about $4\%$ larger, $\left <r \right>=10.48 \pm 0.21 \,kpc$. The determination is tied to the $p=1.35$, having this value adopted from the \cite{NARDETTO2004} modeling of $\delta$ Cep - the prototype classical Cepheid. However, in the original work of \cite{NARDETTO2004}, the authors claim that in the case of broadband interferometric observations, the value of $p=1.27$ should rather be used as it is associated with the pulsation velocity related to the continuum of the photosphere. Assuming a constant $p \varpi$ solution, the corresponding scaling would bring the mean M 3 distance reported in the \cite{JURCSIK0} work to the value of $\left <r \right>=9.86\,kpc$ that is slightly closer to the mean M 3 distance we obtained in this work.

We obtained very precise radii determinations when assuming uniform M 3 distance with the rms of fitted relations corresponding to around 0.5\% of a stellar radius. We observe much better agreement of PR relations with the theoretical prediction of \cite{MARCONI2015} than in the case of determinations based on the Gaia DR3 parallaxes of individual RR Lyrae stars from the solar neighborhood (\citealt{GARANCE}, Z24). We estimated the systematic uncertainty of the determined mean stellar radii of $0.17 R_{\odot}$ with its largest component $\left(0.11 R_{\odot}\right)$ being associated with the assumed cluster distance error of 2\%. In this context, the aforementioned remarkable agreement with the theoretical prediction of the PR relation is a good independent check of the accuracy of our mean cluster distance determination as a slight error in the assumed distance yields a relatively substantial systematic shift of values of the determined radii. Namely, the assumption of the cluster distance larger by, e.g., $0.5\,kpc$ would raise systematically determined radii by $0.27 R_{\odot}$. It would make our empirical PR relation derived in this work less compatible with the relation of \cite{MARCONI2015}. Our Zenodo archive includes PR relations based on $\left< r \right>$ distances larger by $0.5\,kpc$ than determined in this work.

The IRSB technique is demanding observationally, but fully empirical and independent of other distance indicators, including the PL relations for different species of pulsating stars. One of its biggest advantages is the small dependence on the reddening and metallicity. High precision of the used data is of crucial importance. Especially the low scatter of the infrared photometric measurements around the corresponding light curve could allow for the mitigation of the statistical uncertainties of individual stellar distances.
In the future, we aim to utilize the technique in the determinations of distances to more systems, such as globular clusters or dwarf spheroidal galaxies in the vicinity of the Milky Way. In the context of the cosmic distance ladder, the technique applied to RR Lyrae stars and Type II Cepheids might allow determining benchmark distances to older stellar systems and yield an independent calibration of the Tip of the Red Giant Branch (TRGB) method that allows to reach Supernova host galaxies. The biggest range limitation in the case of RR Lyrae stars is the spectroscopy. RR Lyraes in the Local Group galaxies are already faint \footnote{RR Lyraes in the LMC have their magnitudes $V \approx 19 \,mag$.}, which makes it challenging to obtain their RV curves based on precise measurements. However, with the advent of the new instruments with bigger apertures, farther targets will be achievable for the method. Classical Cepheids remain the most important pulsating stellar distance indicators that allow reaching much farther, but the calibration of the Baade-Wesselink method for them is a complicated and open issue \citep{TRAHIN}, at least when it comes to determining distances to individual stars. Besides that, the calibration of distance determination methods based on RR Lyrae stars is a pivotal task for the study of previously mentioned systems where classical Cepheids are not present. We expect to perform new, refined calibrations of the technique based on the updated Gaia parallaxes and more photometric and spectroscopic measurements of different types of pulsating stars from the solar neighborhood in the future.

\section{Data availability}

The collection of plots corresponding to all stars from our sample and all five applied SBCRs is available on Zenodo under the following link: \url{https://zenodo.org/records/17613276}.

\begin{acknowledgements}
The research leading to these results has received funding from the European Research Council (ERC) under the European Union’s Horizon 2020 research and innovation program (grant agreement No. 951549). NN, WG, GP, AG and PK acknowledge the support of the French Agence Nationale de la Recherche (ANR), under grant ANR-23-CE31-0009-01 (Unlock-pfactor). AG acknowledges the support of the Agencia Nacional de Investigación Científica y Desarrollo (ANID) through the FONDECYT Regular grant 1241073. We thank the anonymous referee for their thorough insight that particularly helped us improving the structure and legibility of the manuscript.
\end{acknowledgements}

\begin{table}
\centering
\caption{Coefficients of the $p=a \times \left(\log P +0.25\right)+b $ relation together with their uncertainties depending on a SBCR and resulting from the calibration work (Z24). Mean values of $p$ resulting from calibrations are given in the last column.}
\begin{tabular}{|c|c|c|c|c|c|}
\hline
SBCR & $a$ & $\sigma_a$ & $b$ & $\sigma_b$ & $\left< p\right>$ \\
\hline
\hline 
 \multicolumn{6}{|c|}{BISECTOR} \\
\hline
K2004a & -1.44 & 0.40 & 1.453 & 0.019 & 1.44\\
\hline
K2004b & -1.44 & 0.44 & 1.415 & 0.021 & 1.40\\
\hline
G2021 & -1.06 & 0.47 & 1.398 & 0.022 & 1.39\\
\hline
S2021 & -1.17 & 0.46 & 1.453 & 0.022 & 1.44\\
\hline
B2025 & -1.15 & 0.46 & 1.365 & 0.022 & 1.35\\
\hline
\hline
\multicolumn{6}{|c|}{LEAST SQUARES} \\
\hline
K2004a & -1.30 & 0.58 & 1.388 & 0.027 & 1.38\\
\hline
K2004b & -1.31 & 0.58 & 1.359 & 0.027 & 1.35\\
\hline
G2021 & -0.98 & 0.62 & 1.337 & 0.029 & 1.33\\
\hline
S2021 & -1.08 & 0.64 & 1.385 & 0.030 & 1.38\\
\hline
B2025 & -1.07 & 0.61 & 1.307 & 0.029 & 1.30\\
\hline
\end{tabular}
\tablefoot{\citealt{KER_SBCRI}-K2004a; \citealt{KER_SBCRII}-K2004b; \citealt{GRACZ-SBCR}-G2021; \citealt{SALSI}-S2021; \citealt{SBCR-B2025}-B2025.}
\label{tab:pP}
\end{table}

\begin{table*}
\centering
\caption{Summary of the obtained mean distances to stars analyzed in this work assuming $\log P - p$ relations and constant values of $p$ from Table \ref{tab:pP}. Reported statistical uncertainties are standard errors of the mean of individual distances.}
\begin{tabular}{|c||c|c||c|c|}
\hline
$[kpc]$ & \multicolumn{2}{|c||}{BISECTOR} &  \multicolumn{2}{|c|}{LEAST SQUARES} \\
\hline
 SBCR & $\left< r \right>$ & $\left< r \right> (p=const.) $ & $\left< r \right>$ & $\left< r \right> (p=const.)$ \\
\hline
\hline
\cite{KER_SBCRI} & $10.14 \pm 0.19 $ &  $10.21 \pm 0.24 $ & $9.95 \pm 0.20$ & $10.04 \pm 0.26$\\
\hline
\cite{KER_SBCRII} & $10.09 \pm 0.19 $ &  $10.14 \pm 0.24 $ & $9.95 \pm 0.20$ & $10.04 \pm 0.26$\\
\hline
\cite{GRACZ-SBCR} & $10.04 \pm 0.19 $ &  $10.09 \pm 0.20 $ & $10.08 \pm 0.20$ & $10.13 \pm 0.22$\\
\hline
\cite{SALSI} & $10.06 \pm 0.19 $ &  $10.09 \pm 0.20 $ & $10.04 \pm 0.20$ & $10.11 \pm 0.23$\\
\hline
\cite{SBCR-B2025} & $10.03 \pm 0.19 $ &  $10.04 \pm 0.20 $ & $10.08 \pm 0.20$ & $10.15 \pm 0.22$\\
\hline
\hline
mean $\left< r \right>$ & $10.07$ & $10.11$ & $10.02$ & $10.09$ \\
\hline
$\sigma$ & 0.040 & 0.057 & 0.059 & 0.046 \\
\hline
\end{tabular}
\label{tab:meandist}
\end{table*}

\begin{table*}
\centering
\caption{Coefficients of the $\log \left({\left<R\right>}/R_{\odot}\right)=a \times \left(\log P +0.25\right)+b $ relation together with their uncertainties and the rms scatter around the fitted relation obtained in this work depending on the SBCR.}
\begin{tabular}{|c|c|c|c|c|c|}
\hline
SBCR & $a$ & $\sigma_a$ & $b$ & $\sigma_b$ & $rms$\\
\hline
\hline
\multicolumn{6}{|c|}{BISECTOR} \\
\hline
\cite{KER_SBCRI} & 0.562 & 0.025 & 0.7325 & 0.0014 & 0.0048\\
\hline
\cite{KER_SBCRII} & 0.551 & 0.025 & 0.7390 & 0.0014 & 0.0047\\
\hline
\cite{GRACZ-SBCR} & 0.561 & 0.027 & 0.7220 & 0.0015 & 0.0051\\
\hline
\cite{SALSI} & 0.572 & 0.028 & 0.7224 & 0.0015 & 0.0052\\
\hline
\cite{SBCR-B2025} & 0.552 & 0.026 & 0.7268 & 0.0014 & 0.0050\\
\hline
\hline
\multicolumn{6}{|c|}{LEAST SQUARES} \\
\hline
\cite{KER_SBCRI} & 0.564 & 0.025 & 0.7242 & 0.0014 & 0.0048\\
\hline
\cite{KER_SBCRII} & 0.552 & 0.024 & 0.7328 & 0.0014 & 0.0046\\
\hline
\cite{GRACZ-SBCR} & 0.564 & 0.027 & 0.7236 & 0.0015 & 0.0050\\
\hline
\cite{SALSI} & 0.575 & 0.027 & 0.7213 & 0.0015 & 0.0052\\
\hline
\cite{SBCR-B2025} & 0.555 & 0.027 & 0.7288 & 0.0014 & 0.0049\\
\hline
\end{tabular}
\label{tab:PR}
\end{table*}

\begin{appendix}
\onecolumn
\section{Values of the derived $p \varpi$ product}

\begin{center}
\begin{longtable}{|c|c|c|c|c|c|c|}
\caption{Values of the $p \varpi$ product resulting from bisector fit and LS fit together with phase shifts between different observables determined for all stars analyzed in this work. Designation of the SBCRs: \citealt{KER_SBCRI}-K2004a; \citealt{KER_SBCRII}-K2004b; \citealt{GRACZ-SBCR}-G2021; \citealt{SALSI}-S2021; \citealt{SBCR-B2025}-B2025.} \label{tab:products} \\
\hline
\multirow{2}{*}{OBJ}	& \multirow{2}{*}{$\Delta \phi_{V-K}$} & \multirow{2}{*}{$\Delta \phi_{RV-K}$}	& $p \varpi$ [$\mu as$] & $p \varpi$ [$\mu as$] & \multirow{2}{*}{SBCR} & \multirow{2}{*}{Period [d]} \\
 &  & &BISECTOR & LEAST SQUARES & &  \\
\hline
\endfirsthead

\multicolumn{3}{c}
{{\bfseries \tablename\ \thetable{} -- continued from previous page}} \\
\hline
\multirow{2}{*}{OBJ}	& \multirow{2}{*}{$\Delta \phi_{V-K}$} & \multirow{2}{*}{$\Delta \phi_{RV-K}$}	& $p \varpi$ [$\mu as$] & $p \varpi$ [$\mu as$] & \multirow{2}{*}{SBCR} & \multirow{2}{*}{Period [d]} \\
 &  & &BISECTOR & LEAST SQUARES & &  \\
\hline
\endhead

\hline \multicolumn{7}{|r|}{{Continued on next page}} \\ \hline
\endfoot

\hline \hline
\endlastfoot
\hline
\hline
\multirow{5}{*}{V19}& -0.004 & -0.004 &$ 120.066 \pm 0.087 $&$ 114.5 \pm 8.4 $& K2004a & \multirow{5}{*}{0.6319782}\\
\cline{2-6}
& 0.006 & -0.006 &$ 117.124 \pm 0.083 $&$ 111.7 \pm 8.2 $& K2004b & \\
\cline{2-6}
& -0.006 & 0.01 &$ 126.65 \pm 0.15 $&$ 117.0 \pm 11.0 $& G2021 & \\
\cline{2-6}
& -0.012 & 0.008 &$ 128.94 \pm 0.16 $&$ 119.0 \pm 11.0 $& S2021 & \\
\cline{2-6}
& 0.004 & 0.01 &$ 122.78 \pm 0.14 $&$ 114.0 \pm 11.0 $& B2025 & \\
\hline
\hline
\multirow{5}{*}{V26}& 0.004 & 0.036 &$ 155.114 \pm 0.047 $&$ 152.9 \pm 6.2 $& K2004a & \multirow{5}{*}{0.597744}\\
\cline{2-6}
& 0.014 & 0.038 &$ 152.141 \pm 0.045 $&$ 149.9 \pm 6.1 $& K2004b & \\
\cline{2-6}
& 0.01 & 0.04 &$ 151.849 \pm 0.089 $&$ 147.8 \pm 8.2 $& G2021 & \\
\cline{2-6}
& 0.0 & 0.038 &$ 157.207 \pm 0.089 $&$ 153.2 \pm 8.2 $& S2021 & \\
\cline{2-6}
& 0.016 & 0.04 &$ 147.745 \pm 0.083 $&$ 143.8 \pm 7.9 $& B2025 & \\
\hline
\hline
\multirow{5}{*}{V55}& -0.13 & -0.082 &$ 148.946 \pm 0.05 $&$ 147.6 \pm 4.7 $& K2004a & \multirow{5}{*}{0.5298165}\\
\cline{2-6}
& -0.12 & -0.08 &$ 144.759 \pm 0.057 $&$ 143.3 \pm 4.8 $& K2004b & \\
\cline{2-6}
& -0.118 & -0.084 &$ 144.29 \pm 0.15 $&$ 140.9 \pm 7.2 $& G2021 & \\
\cline{2-6}
& -0.13 & -0.084 &$ 151.61 \pm 0.13 $&$ 148.6 \pm 7.1 $& S2021 & \\
\cline{2-6}
& -0.11 & -0.082 &$ 142.12 \pm 0.15 $&$ 138.6 \pm 7.3 $& B2025 & \\
\hline
\hline
\multirow{5}{*}{V36}& -0.068 & -0.034 &$ 140.611 \pm 0.058 $&$ 139.1 \pm 4.8 $& K2004a & \multirow{5}{*}{0.54559}\\
\cline{2-6}
& -0.058 & -0.034 &$ 136.279 \pm 0.055 $&$ 134.8 \pm 4.7 $& K2004b & \\
\cline{2-6}
& -0.062 & -0.04 &$ 126.5 \pm 0.1 $&$ 123.7 \pm 6.3 $& G2021 & \\
\cline{2-6}
& -0.074 & -0.04 &$ 133.69 \pm 0.11 $&$ 130.7 \pm 6.6 $& S2021 & \\
\cline{2-6}
& -0.054 & -0.04 &$ 124.538 \pm 0.1 $&$ 121.7 \pm 6.2 $& B2025 & \\
\hline
\hline
\multirow{5}{*}{V84}& -0.002 & 0.014 &$ 151.287 \pm 0.086 $&$ 148.3 \pm 7.0 $& K2004a & \multirow{5}{*}{0.59572885}\\
\cline{2-6}
& 0.006 & 0.014 &$ 147.477 \pm 0.079 $&$ 144.5 \pm 7.0 $& K2004b & \\
\cline{2-6}
& 0.0 & 0.016 &$ 153.84 \pm 0.14 $&$ 148.7 \pm 9.2 $& G2021 & \\
\cline{2-6}
& -0.006 & 0.016 &$ 159.62 \pm 0.15 $&$ 154.6 \pm 9.3 $& S2021 & \\
\cline{2-6}
& 0.006 & 0.016 &$ 149.76 \pm 0.13 $&$ 144.8 \pm 9.0 $& B2025 & \\
\hline
\hline
\multirow{5}{*}{V89}& -0.044 & 0.006 &$ 147.166 \pm 0.098 $&$ 144.3 \pm 6.7 $& K2004a & \multirow{5}{*}{0.5484783}\\
\cline{2-6}
& -0.034 & 0.008 &$ 143.2 \pm 0.1 $&$ 140.4 \pm 6.6 $& K2004b & \\
\cline{2-6}
& -0.03 & 0.008 &$ 140.15 \pm 0.27 $&$ 133.9 \pm 9.6 $& G2021 & \\
\cline{2-6}
& -0.042 & 0.006 &$ 146.19 \pm 0.26 $&$ 140.0 \pm 9.8 $& S2021 & \\
\cline{2-6}
& -0.022 & 0.008 &$ 136.14 \pm 0.26 $&$ 130.1 \pm 9.4 $& B2025 & \\
\hline
\hline
\multirow{5}{*}{V60}& -0.032 & -0.01 &$ 132.764 \pm 0.099 $&$ 128.0 \pm 8.3 $& K2004a & \multirow{5}{*}{0.707725}\\
\cline{2-6}
& -0.024 & -0.01 &$ 130.166 \pm 0.096 $&$ 125.6 \pm 8.0 $& K2004b & \\
\cline{2-6}
& -0.034 & -0.014 &$ 127.01 \pm 0.15 $&$ 119.0 \pm 10.0 $& G2021 & \\
\cline{2-6}
& -0.04 & -0.014 &$ 130.93 \pm 0.16 $&$ 123.0 \pm 11.0 $& S2021 & \\
\cline{2-6}
& -0.024 & -0.012 &$ 124.87 \pm 0.13 $&$ 117.0 \pm 10.0 $& B2025 & \\
\hline
\hline
\multirow{5}{*}{V72}& -0.092 & -0.052 &$ 153.229 \pm 0.09 $&$ 149.1 \pm 8.3 $& K2004a & \multirow{5}{*}{0.4560759}\\
\cline{2-6}
& -0.078 & -0.048 &$ 150.179 \pm 0.093 $&$ 145.9 \pm 8.4 $& K2004b & \\
\cline{2-6}
& -0.078 & -0.05 &$ 145.77 \pm 0.18 $&$ 136.0 \pm 12.0 $& G2021 & \\
\cline{2-6}
& -0.092 & -0.052 &$ 152.32 \pm 0.17 $&$ 143.0 \pm 12.0 $& S2021 & \\
\cline{2-6}
& -0.068 & -0.05 &$ 141.96 \pm 0.19 $&$ 132.0 \pm 12.0 $& B2025 & \\
\hline
\hline
\multirow{5}{*}{V124}& -0.06 & -0.018 &$ 132.67 \pm 0.16 $&$ 127.6 \pm 8.5 $& K2004a & \multirow{5}{*}{0.752436}\\
\cline{2-6}
& -0.052 & -0.012 &$ 131.96 \pm 0.15 $&$ 127.0 \pm 8.4 $& K2004b & \\
\cline{2-6}
& -0.05 & 0.0 &$ 143.91 \pm 0.26 $&$ 136.0 \pm 11.0 $& G2021 & \\
\cline{2-6}
& -0.054 & -0.002 &$ 147.08 \pm 0.26 $&$ 139.0 \pm 11.0 $& S2021 & \\
\cline{2-6}
& -0.044 & 0.0 &$ 138.9 \pm 0.25 $&$ 131.0 \pm 11.0 $& B2025 & \\
\hline
\hline
\multirow{5}{*}{V46}& -0.116 & -0.1 &$ 125.885 \pm 0.071 $&$ 122.1 \pm 7.1 $& K2004a & \multirow{5}{*}{0.6133736}\\
\cline{2-6}
& -0.108 & -0.1 &$ 123.48 \pm 0.071 $&$ 119.6 \pm 7.2 $& K2004b & \\
\cline{2-6}
& -0.122 & -0.096 &$ 126.22 \pm 0.14 $&$ 119.7 \pm 9.4 $& G2021 & \\
\cline{2-6}
& -0.126 & -0.098 &$ 128.87 \pm 0.14 $&$ 122.3 \pm 9.5 $& S2021 & \\
\cline{2-6}
& -0.112 & -0.096 &$ 122.88 \pm 0.14 $&$ 116.2 \pm 9.3 $& B2025 & \\
\hline
\hline
\pagebreak
\multirow{5}{*}{V27}& 0.158 & 0.174 &$ 128.268 \pm 0.054 $&$ 125.5 \pm 6.2 $& K2004a & \multirow{5}{*}{0.579083}\\
\cline{2-6}
& 0.168 & 0.174 &$ 125.268 \pm 0.052 $&$ 122.6 \pm 6.1 $& K2004b & \\
\cline{2-6}
& 0.16 & 0.168 &$ 126.562 \pm 0.089 $&$ 121.9 \pm 8.0 $& G2021 & \\
\cline{2-6}
& 0.152 & 0.168 &$ 131.494 \pm 0.098 $&$ 126.7 \pm 8.2 $& S2021 & \\
\cline{2-6}
& 0.17 & 0.166 &$ 123.298 \pm 0.093 $&$ 118.6 \pm 7.9 $& B2025 & \\
\hline
\hline
\multirow{5}{*}{V51}& 0.18 & 0.092 &$ 137.51 \pm 0.1 $&$ 133.6 \pm 7.6 $& K2004a & \multirow{5}{*}{0.5839775}\\
\cline{2-6}
& 0.19 & 0.094 &$ 134.9 \pm 0.098 $&$ 130.9 \pm 7.6 $& K2004b & \\
\cline{2-6}
& 0.182 & 0.096 &$ 134.18 \pm 0.18 $&$ 127.0 \pm 10.0 $& G2021 & \\
\cline{2-6}
& 0.174 & 0.096 &$ 139.79 \pm 0.2 $&$ 133.0 \pm 10.0 $& S2021 & \\
\cline{2-6}
& 0.192 & 0.098 &$ 131.59 \pm 0.16 $&$ 124.0 \pm 10.0 $& B2025 & \\
\hline
\hline
\multirow{5}{*}{V81}& -0.18 & -0.052 &$ 152.914 \pm 0.098 $&$ 149.4 \pm 7.6 $& K2004a & \multirow{5}{*}{0.5291139}\\
\cline{2-6}
& -0.168 & -0.05 &$ 151.136 \pm 0.096 $&$ 147.7 \pm 7.5 $& K2004b & \\
\cline{2-6}
& -0.17 & -0.054 &$ 139.45 \pm 0.11 $&$ 133.5 \pm 9.4 $& G2021 & \\
\cline{2-6}
& -0.182 & -0.054 &$ 145.53 \pm 0.13 $&$ 139.7 \pm 9.5 $& S2021 & \\
\cline{2-6}
& -0.162 & -0.054 &$ 136.58 \pm 0.13 $&$ 130.6 \pm 9.3 $& B2025 & \\
\hline
\hline
\multirow{5}{*}{V83}& -0.28 & -0.224 &$ 162.87 \pm 0.054 $&$ 159.6 \pm 7.6 $& K2004a & \multirow{5}{*}{0.5012495}\\
\cline{2-6}
& -0.268 & -0.222 &$ 159.934 \pm 0.056 $&$ 156.5 \pm 7.8 $& K2004b & \\
\cline{2-6}
& -0.278 & -0.226 &$ 151.89 \pm 0.1 $&$ 145.0 \pm 10.0 $& G2021 & \\
\cline{2-6}
& -0.29 & -0.228 &$ 156.495 \pm 0.092 $&$ 150.9 \pm 9.7 $& S2021 & \\
\cline{2-6}
& -0.268 & -0.224 &$ 150.37 \pm 0.11 $&$ 143.0 \pm 11.0 $& B2025 & \\
\hline
\end{longtable}
\end{center}

\section{Individual stellar distances}

\begin{center}
\begin{longtable}{|c||c|c|c|c|c||c|c|c|c|c||c|}
\caption{Individual stellar distances together with the mean distances and the corresponding statistical uncertainties depending on a SBCR obtained in this work. Designation of the SBCRs: \citealt{KER_SBCRI}-K2004a; \citealt{KER_SBCRII}-K2004b; \citealt{GRACZ-SBCR}-G2021; \citealt{SALSI}-S2021; \citealt{SBCR-B2025}-B2025.} \label{tab:dist} \\
\hline
\multirow{3}{*}{OBJ} & \multicolumn{5}{c||}{BISECTOR}  & \multicolumn{5}{c||}{LEAST SQUARES} & \\
\cline{2-11}
 & \multirow{2}{*}{$r\,[kpc]$} & \multirow{2}{*}{rms ($r$)} & \multicolumn{3}{c||}{percentiles} & \multirow{2}{*}{$r\,[kpc]$} & \multirow{2}{*}{rms ($r$)} & \multicolumn{3}{c||}{percentiles} & SBCR \\
\cline{4-6}
\cline{9-11}
 & & &15.85 & 50.00 &84.15 & & & 15.85 & 50.00 &84.15 &  \\
\hline
\hline
\endfirsthead

\multicolumn{3}{c}
{{\bfseries \tablename\ \thetable{} -- continued from previous page}} \\
\hline
\multirow{3}{*}{OBJ} & \multicolumn{5}{c||}{BISECTOR}  & \multicolumn{5}{c||}{LEAST SQUARES} & \multirow{3}{*}{SBCR}\\
\cline{2-11}
 & \multirow{2}{*}{$r\,[kpc]$} & \multirow{2}{*}{rms ($r$)} & \multicolumn{3}{c||}{percentiles} & \multirow{2}{*}{$r\,[kpc]$} & \multirow{2}{*}{rms ($r$)} & \multicolumn{3}{c||}{percentiles} & \\
\cline{4-6}
\cline{9-11}
 & & &15.85 & 50.00 &84.15 & & & 15.85 & 50.00 &84.15 &  \\
\hline
\hline
\endhead

\hline \multicolumn{12}{|r|}{{Continued on next page}} \\ \hline
\endfoot

\hline \hline
\endlastfoot

\multirow{5}{*}{V19}& 11.49 & 0.86 & 10.1 & 10.88 & 11.8 & 11.55 & 1.04 & 10.58 & 11.54 & 12.62 & K2004a \\
\cline{2-12}
& 11.46 & 0.84 & 10.07 & 10.82 & 11.69 & 11.57 & 1.06 & 10.62 & 11.57 & 12.68 & K2004b \\
\cline{2-12}
& 10.61 & 0.87 & 8.97 & 9.78 & 10.7 & 11.0 & 1.26 & 9.89 & 11.0 & 12.33 & G2021 \\
\cline{2-12}
& 10.81 & 0.9 & 9.11 & 9.92 & 10.89 & 11.18 & 1.27 & 10.04 & 11.19 & 12.53 & S2021 \\
\cline{2-12}
& 10.64 & 0.88 & 8.98 & 9.81 & 10.73 & 10.99 & 1.26 & 9.89 & 11.04 & 12.35 & B2025 \\
\hline
\hline
\multirow{5}{*}{V26}& 9.12 & 0.36 & 8.63 & 9.0 & 9.37 & 8.85 & 0.4 & 8.47 & 8.84 & 9.25 & K2004a \\
\cline{2-12}
& 9.05 & 0.37 & 8.58 & 8.93 & 9.32 & 8.83 & 0.4 & 8.44 & 8.82 & 9.23 & K2004b \\
\cline{2-12}
& 9.02 & 0.46 & 8.38 & 8.83 & 9.28 & 8.87 & 0.5 & 8.4 & 8.89 & 9.41 & G2021 \\
\cline{2-12}
& 9.05 & 0.44 & 8.42 & 8.85 & 9.31 & 8.85 & 0.5 & 8.35 & 8.84 & 9.37 & S2021 \\
\cline{2-12}
& 9.03 & 0.46 & 8.38 & 8.82 & 9.29 & 8.89 & 0.52 & 8.41 & 8.9 & 9.43 & B2025 \\
\hline
\hline
\multirow{5}{*}{V55}& 10.01 & 0.45 & 9.41 & 9.82 & 10.29 & 9.63 & 0.47 & 9.17 & 9.63 & 10.12 & K2004a \\
\cline{2-12}
& 10.03 & 0.46 & 9.42 & 9.84 & 10.32 & 9.72 & 0.49 & 9.26 & 9.71 & 10.25 & K2004b \\
\cline{2-12}
& 9.88 & 0.57 & 9.01 & 9.57 & 10.16 & 9.67 & 0.65 & 9.06 & 9.66 & 10.36 & G2021 \\
\cline{2-12}
& 9.78 & 0.56 & 8.96 & 9.52 & 10.1 & 9.51 & 0.61 & 8.96 & 9.52 & 10.16 & S2021 \\
\cline{2-12}
& 9.81 & 0.57 & 8.94 & 9.49 & 10.1 & 9.63 & 0.67 & 9.01 & 9.63 & 10.3 & B2025 \\
\hline
\hline
\multirow{5}{*}{V36}& 10.47 & 0.47 & 9.83 & 10.27 & 10.79 & 10.1 & 0.51 & 9.62 & 10.1 & 10.62 & K2004a \\
\cline{2-12}
& 10.52 & 0.48 & 9.85 & 10.32 & 10.81 & 10.21 & 0.51 & 9.72 & 10.2 & 10.73 & K2004b \\
\cline{2-12}
& 11.16 & 0.7 & 10.09 & 10.77 & 11.47 & 10.91 & 0.81 & 10.17 & 10.9 & 11.76 & G2021 \\
\cline{2-12}
& 10.98 & 0.67 & 9.96 & 10.57 & 11.3 & 10.71 & 0.76 & 10.03 & 10.71 & 11.53 & S2021 \\
\cline{2-12}
& 11.08 & 0.71 & 10.0 & 10.65 & 11.41 & 10.85 & 0.78 & 10.15 & 10.87 & 11.69 & B2025 \\
\hline
\hline
\multirow{5}{*}{V84}& 9.37 & 0.42 & 8.81 & 9.2 & 9.64 & 9.14 & 0.45 & 8.72 & 9.15 & 9.61 & K2004a \\
\cline{2-12}
& 9.35 & 0.43 & 8.76 & 9.18 & 9.62 & 9.18 & 0.46 & 8.74 & 9.18 & 9.66 & K2004b \\
\cline{2-12}
& 8.91 & 0.49 & 8.2 & 8.66 & 9.18 & 8.83 & 0.56 & 8.3 & 8.82 & 9.42 & G2021 \\
\cline{2-12}
& 8.92 & 0.49 & 8.22 & 8.69 & 9.19 & 8.78 & 0.55 & 8.28 & 8.79 & 9.37 & S2021 \\
\cline{2-12}
& 8.92 & 0.5 & 8.17 & 8.67 & 9.18 & 8.84 & 0.56 & 8.33 & 8.82 & 9.43 & B2025 \\
\hline
\hline
\multirow{5}{*}{V89}& 9.98 & 0.5 & 9.26 & 9.73 & 10.25 & 9.72 & 0.54 & 9.2 & 9.72 & 10.29 & K2004a \\
\cline{2-12}
& 9.99 & 0.5 & 9.27 & 9.75 & 10.28 & 9.78 & 0.55 & 9.26 & 9.77 & 10.33 & K2004b \\
\cline{2-12}
& 10.06 & 0.64 & 9.04 & 9.63 & 10.3 & 10.06 & 0.76 & 9.34 & 10.07 & 10.88 & G2021 \\
\cline{2-12}
& 10.03 & 0.63 & 9.02 & 9.6 & 10.27 & 9.98 & 0.75 & 9.31 & 9.98 & 10.77 & S2021 \\
\cline{2-12}
& 10.12 & 0.65 & 9.07 & 9.68 & 10.35 & 10.14 & 0.77 & 9.43 & 10.15 & 10.97 & B2025 \\
\hline
\hline
\multirow{5}{*}{V60}& 9.86 & 0.67 & 8.85 & 9.47 & 10.17 & 9.83 & 0.85 & 9.03 & 9.85 & 10.72 & K2004a \\
\cline{2-12}
& 9.77 & 0.67 & 8.75 & 9.39 & 10.08 & 9.78 & 0.85 & 8.95 & 9.76 & 10.64 & K2004b \\
\cline{2-12}
& 10.17 & 0.86 & 8.72 & 9.51 & 10.4 & 10.41 & 1.15 & 9.38 & 10.41 & 11.65 & G2021 \\
\cline{2-12}
& 10.21 & 0.84 & 8.76 & 9.51 & 10.4 & 10.38 & 1.16 & 9.42 & 10.43 & 11.66 & S2021 \\
\cline{2-12}
& 10.01 & 0.85 & 8.57 & 9.36 & 10.26 & 10.26 & 1.1 & 9.22 & 10.24 & 11.38 & B2025 \\
\hline
\hline
\multirow{5}{*}{V72}& 10.34 & 0.7 & 9.21 & 9.89 & 10.6 & 10.1 & 0.84 & 9.3 & 10.09 & 10.98 & K2004a \\
\cline{2-12}
& 10.29 & 0.72 & 9.14 & 9.81 & 10.55 & 10.13 & 0.85 & 9.33 & 10.13 & 10.99 & K2004b \\
\cline{2-12}
& 10.25 & 0.9 & 8.56 & 9.34 & 10.32 & 10.49 & 1.24 & 9.39 & 10.48 & 11.8 & G2021 \\
\cline{2-12}
& 10.24 & 0.86 & 8.63 & 9.42 & 10.35 & 10.37 & 1.2 & 9.29 & 10.36 & 11.67 & S2021 \\
\cline{2-12}
& 10.35 & 0.88 & 8.64 & 9.44 & 10.39 & 10.64 & 1.27 & 9.51 & 10.64 & 11.99 & B2025 \\
\hline
\hline
\multirow{5}{*}{V124}& 9.58 & 0.7 & 8.57 & 9.23 & 9.94 & 9.59 & 0.91 & 8.74 & 9.6 & 10.52 & K2004a \\
\cline{2-12}
& 9.34 & 0.7 & 8.32 & 8.97 & 9.7 & 9.4 & 0.89 & 8.56 & 9.41 & 10.34 & K2004b \\
\cline{2-12}
& 8.78 & 0.73 & 7.61 & 8.31 & 9.07 & 8.92 & 0.97 & 8.01 & 8.89 & 9.92 & G2021 \\
\cline{2-12}
& 8.87 & 0.73 & 7.69 & 8.38 & 9.14 & 8.98 & 0.99 & 8.09 & 9.0 & 10.06 & S2021 \\
\cline{2-12}
& 8.78 & 0.74 & 7.57 & 8.27 & 9.03 & 8.94 & 1.01 & 8.01 & 8.93 & 9.98 & B2025 \\
\hline
\hline
\multirow{5}{*}{V46}& 11.11 & 0.69 & 10.06 & 10.71 & 11.39 & 10.97 & 0.76 & 10.25 & 10.97 & 11.75 & K2004a \\
\cline{2-12}
& 11.02 & 0.69 & 9.94 & 10.6 & 11.31 & 10.95 & 0.8 & 10.21 & 10.95 & 11.8 & K2004b \\
\cline{2-12}
& 10.76 & 0.8 & 9.37 & 10.1 & 10.93 & 10.86 & 1.02 & 9.94 & 10.84 & 11.93 & G2021 \\
\cline{2-12}
& 10.93 & 0.8 & 9.55 & 10.26 & 11.12 & 10.99 & 1.03 & 10.05 & 10.97 & 12.07 & S2021 \\
\cline{2-12}
& 10.76 & 0.8 & 9.32 & 10.06 & 10.88 & 10.9 & 1.04 & 9.94 & 10.9 & 11.99 & B2025 \\
\hline
\hline
\multirow{5}{*}{V27}& 11.18 & 0.55 & 10.4 & 10.95 & 11.51 & 10.93 & 0.59 & 10.37 & 10.92 & 11.54 & K2004a \\
\cline{2-12}
& 11.15 & 0.55 & 10.35 & 10.87 & 11.46 & 10.95 & 0.61 & 10.37 & 10.95 & 11.59 & K2004b \\
\cline{2-12}
& 10.94 & 0.68 & 9.86 & 10.51 & 11.22 & 10.87 & 0.8 & 10.12 & 10.87 & 11.72 & G2021 \\
\cline{2-12}
& 10.94 & 0.68 & 9.88 & 10.52 & 11.24 & 10.82 & 0.78 & 10.09 & 10.78 & 11.63 & S2021 \\
\cline{2-12}
& 10.95 & 0.7 & 9.86 & 10.51 & 11.26 & 10.91 & 0.81 & 10.19 & 10.91 & 11.79 & B2025 \\
\hline
\hline
\multirow{5}{*}{V51}& 10.39 & 0.53 & 9.61 & 10.12 & 10.66 & 10.23 & 0.59 & 9.66 & 10.23 & 10.84 & K2004a \\
\cline{2-12}
& 10.31 & 0.53 & 9.57 & 10.05 & 10.62 & 10.22 & 0.6 & 9.67 & 10.22 & 10.85 & K2004b \\
\cline{2-12}
& 10.29 & 0.67 & 9.21 & 9.83 & 10.54 & 10.4 & 0.81 & 9.67 & 10.4 & 11.25 & G2021 \\
\cline{2-12}
& 10.26 & 0.66 & 9.2 & 9.82 & 10.5 & 10.28 & 0.78 & 9.58 & 10.3 & 11.12 & S2021 \\
\cline{2-12}
& 10.23 & 0.66 & 9.15 & 9.76 & 10.45 & 10.4 & 0.81 & 9.63 & 10.37 & 11.22 & B2025 \\
\hline
\hline
\multirow{5}{*}{V81}& 9.75 & 0.64 & 8.72 & 9.33 & 10.01 & 9.52 & 0.72 & 8.86 & 9.52 & 10.28 & K2004a \\
\cline{2-12}
& 9.61 & 0.64 & 8.58 & 9.18 & 9.84 & 9.44 & 0.72 & 8.77 & 9.43 & 10.16 & K2004b \\
\cline{2-12}
& 10.23 & 0.88 & 8.59 & 9.39 & 10.32 & 10.21 & 1.14 & 9.18 & 10.16 & 11.41 & G2021 \\
\cline{2-12}
& 10.2 & 0.85 & 8.66 & 9.42 & 10.31 & 10.12 & 1.09 & 9.16 & 10.11 & 11.25 & S2021 \\
\cline{2-12}
& 10.22 & 0.87 & 8.57 & 9.34 & 10.28 & 10.22 & 1.15 & 9.22 & 10.23 & 11.45 & B2025 \\
\hline
\hline
\multirow{5}{*}{V83}& 9.36 & 0.49 & 8.69 & 9.17 & 9.66 & 9.1 & 0.54 & 8.59 & 9.1 & 9.66 & K2004a \\
\cline{2-12}
& 9.3 & 0.51 & 8.58 & 9.08 & 9.61 & 9.1 & 0.57 & 8.56 & 9.09 & 9.69 & K2004b \\
\cline{2-12}
& 9.55 & 0.66 & 8.54 & 9.13 & 9.84 & 9.56 & 0.8 & 8.81 & 9.52 & 10.36 & G2021 \\
\cline{2-12}
& 9.66 & 0.64 & 8.66 & 9.28 & 9.95 & 9.54 & 0.75 & 8.86 & 9.55 & 10.34 & S2021 \\
\cline{2-12}
& 9.46 & 0.67 & 8.4 & 9.02 & 9.74 & 9.51 & 0.81 & 8.74 & 9.5 & 10.35 & B2025 \\
\hline
\hline

\end{longtable}
\end{center}

\section{Mean stellar radii}

\begin{center}
\begin{longtable}{|c||c|c||c|c||c|}
\caption{Mean stellar radii and their statistical uncertainties obtained in this work.} \label{tab:radii} \\
\hline
\multirow{2}{*}{OBJ} & \multicolumn{2}{c||}{BISECTOR}  & \multicolumn{2}{c||}{LEAST SQUARES} & \multirow{2}{*}{SBCR} \\
\cline{2-5}
 & $R [R_{\odot}]$ & $\sigma_R [R_{\odot}]$ & $R [R_{\odot}]$ & $\sigma_R [R_{\odot}]$ & \\
\hline
\hline
\endfirsthead

\multicolumn{6}{c}
{{\bfseries \tablename\ \thetable{} -- continued from previous page}} \\
\hline
\multirow{2}{*}{OBJ} & \multicolumn{2}{c||}{BISECTOR}  & \multicolumn{2}{c||}{LEAST SQUARES} & \multirow{2}{*}{SBCR} \\
\cline{2-5}
 & $R [R_{\odot}]$ & $\sigma_R [R_{\odot}]$ & $R [R_{\odot}]$ & $\sigma_R [R_{\odot}]$ & \\
\hline
\hline
\endhead

\hline \multicolumn{6}{|r|}{{Continued on next page}} \\ \hline
\endfoot

\hline \hline
\endlastfoot

\multirow{5}{*}{V19}& 5.829 & 0.016 & 5.719 & 0.015 & \citealt{KER_SBCRI} \\
\cline{2-6}
& 5.902 & 0.015 & 5.819 & 0.016 & \citealt{KER_SBCRII} \\
\cline{2-6}
& 5.686 & 0.02 & 5.707 & 0.02 & \citealt{GRACZ-SBCR} \\
\cline{2-6}
& 5.705 & 0.021 & 5.692 & 0.021 & \citealt{SALSI} \\
\cline{2-6}
& 5.737 & 0.02 & 5.764 & 0.02 & \citealt{SBCR-B2025} \\
\hline
\hline
\multirow{5}{*}{V26}& 5.655 & 0.013 & 5.549 & 0.013 & \citealt{KER_SBCRI} \\
\cline{2-6}
& 5.739 & 0.013 & 5.659 & 0.013 & \citealt{KER_SBCRII} \\
\cline{2-6}
& 5.523 & 0.017 & 5.545 & 0.018 & \citealt{GRACZ-SBCR} \\
\cline{2-6}
& 5.528 & 0.017 & 5.518 & 0.018 & \citealt{SALSI} \\
\cline{2-6}
& 5.582 & 0.017 & 5.61 & 0.017 & \citealt{SBCR-B2025} \\
\hline
\hline
\multirow{5}{*}{V55}& 5.189 & 0.016 & 5.091 & 0.016 & \citealt{KER_SBCRI} \\
\cline{2-6}
& 5.271 & 0.016 & 5.197 & 0.015 & \citealt{KER_SBCRII} \\
\cline{2-6}
& 5.086 & 0.02 & 5.103 & 0.02 & \citealt{GRACZ-SBCR} \\
\cline{2-6}
& 5.083 & 0.02 & 5.07 & 0.019 & \citealt{SALSI} \\
\cline{2-6}
& 5.144 & 0.02 & 5.166 & 0.019 & \citealt{SBCR-B2025} \\
\hline
\hline
\multirow{5}{*}{V36}& 5.229 & 0.014 & 5.13 & 0.013 & \citealt{KER_SBCRI} \\
\cline{2-6}
& 5.313 & 0.014 & 5.238 & 0.013 & \citealt{KER_SBCRII} \\
\cline{2-6}
& 5.096 & 0.018 & 5.113 & 0.018 & \citealt{GRACZ-SBCR} \\
\cline{2-6}
& 5.097 & 0.018 & 5.084 & 0.018 & \citealt{SALSI} \\
\cline{2-6}
& 5.157 & 0.017 & 5.18 & 0.017 & \citealt{SBCR-B2025} \\
\hline
\hline
\multirow{5}{*}{V84}& 5.536 & 0.011 & 5.433 & 0.01 & \citealt{KER_SBCRI} \\
\cline{2-6}
& 5.616 & 0.01 & 5.539 & 0.01 & \citealt{KER_SBCRII} \\
\cline{2-6}
& 5.39 & 0.014 & 5.413 & 0.014 & \citealt{GRACZ-SBCR} \\
\cline{2-6}
& 5.399 & 0.014 & 5.389 & 0.014 & \citealt{SALSI} \\
\cline{2-6}
& 5.448 & 0.013 & 5.477 & 0.014 & \citealt{SBCR-B2025} \\
\hline
\hline
\multirow{5}{*}{V89}& 5.281 & 0.014 & 5.183 & 0.013 & \citealt{KER_SBCRI} \\
\cline{2-6}
& 5.366 & 0.014 & 5.292 & 0.014 & \citealt{KER_SBCRII} \\
\cline{2-6}
& 5.154 & 0.018 & 5.175 & 0.018 & \citealt{GRACZ-SBCR} \\
\cline{2-6}
& 5.151 & 0.019 & 5.142 & 0.019 & \citealt{SALSI} \\
\cline{2-6}
& 5.215 & 0.018 & 5.241 & 0.018 & \citealt{SBCR-B2025} \\
\hline
\hline
\multirow{5}{*}{V60}& 6.123 & 0.016 & 6.008 & 0.016 & \citealt{KER_SBCRI} \\
\cline{2-6}
& 6.205 & 0.016 & 6.118 & 0.015 & \citealt{KER_SBCRII} \\
\cline{2-6}
& 5.984 & 0.02 & 6.007 & 0.02 & \citealt{GRACZ-SBCR} \\
\cline{2-6}
& 6.0 & 0.021 & 5.987 & 0.021 & \citealt{SALSI} \\
\cline{2-6}
& 6.042 & 0.02 & 6.071 & 0.02 & \citealt{SBCR-B2025} \\
\hline
\hline
\multirow{5}{*}{V72}& 4.767 & 0.02 & 4.677 & 0.019 & \citealt{KER_SBCRI} \\
\cline{2-6}
& 4.849 & 0.019 & 4.782 & 0.019 & \citealt{KER_SBCRII} \\
\cline{2-6}
& 4.65 & 0.026 & 4.667 & 0.025 & \citealt{GRACZ-SBCR} \\
\cline{2-6}
& 4.645 & 0.026 & 4.634 & 0.026 & \citealt{SALSI} \\
\cline{2-6}
& 4.708 & 0.026 & 4.731 & 0.026 & \citealt{SBCR-B2025} \\
\hline
\hline
\multirow{5}{*}{V124}& 6.284 & 0.013 & 6.167 & 0.013 & \citealt{KER_SBCRI} \\
\cline{2-6}
& 6.361 & 0.013 & 6.273 & 0.013 & \citealt{KER_SBCRII} \\
\cline{2-6}
& 6.133 & 0.017 & 6.158 & 0.017 & \citealt{GRACZ-SBCR} \\
\cline{2-6}
& 6.154 & 0.017 & 6.143 & 0.017 & \citealt{SALSI} \\
\cline{2-6}
& 6.186 & 0.017 & 6.217 & 0.016 & \citealt{SBCR-B2025} \\
\hline
\hline
\multirow{5}{*}{V46}& 5.689 & 0.014 & 5.583 & 0.014 & \citealt{KER_SBCRI} \\
\cline{2-6}
& 5.764 & 0.013 & 5.684 & 0.014 & \citealt{KER_SBCRII} \\
\cline{2-6}
& 5.549 & 0.018 & 5.573 & 0.019 & \citealt{GRACZ-SBCR} \\
\cline{2-6}
& 5.565 & 0.019 & 5.555 & 0.019 & \citealt{SALSI} \\
\cline{2-6}
& 5.603 & 0.018 & 5.632 & 0.018 & \citealt{SBCR-B2025} \\
\hline
\hline
\multirow{5}{*}{V27}& 5.574 & 0.013 & 5.47 & 0.012 & \citealt{KER_SBCRI} \\
\cline{2-6}
& 5.654 & 0.013 & 5.576 & 0.012 & \citealt{KER_SBCRII} \\
\cline{2-6}
& 5.441 & 0.017 & 5.463 & 0.016 & \citealt{GRACZ-SBCR} \\
\cline{2-6}
& 5.449 & 0.017 & 5.439 & 0.016 & \citealt{SALSI} \\
\cline{2-6}
& 5.497 & 0.016 & 5.525 & 0.016 & \citealt{SBCR-B2025} \\
\hline
\hline
\pagebreak
\multirow{5}{*}{V51}& 5.549 & 0.014 & 5.445 & 0.014 & \citealt{KER_SBCRI} \\
\cline{2-6}
& 5.627 & 0.013 & 5.549 & 0.013 & \citealt{KER_SBCRII} \\
\cline{2-6}
& 5.415 & 0.018 & 5.437 & 0.018 & \citealt{GRACZ-SBCR} \\
\cline{2-6}
& 5.425 & 0.018 & 5.415 & 0.018 & \citealt{SALSI} \\
\cline{2-6}
& 5.47 & 0.018 & 5.498 & 0.018 & \citealt{SBCR-B2025} \\
\hline
\hline
\multirow{5}{*}{V81}& 5.34 & 0.022 & 5.236 & 0.021 & \citealt{KER_SBCRI} \\
\cline{2-6}
& 5.422 & 0.022 & 5.343 & 0.022 & \citealt{KER_SBCRII} \\
\cline{2-6}
& 5.225 & 0.027 & 5.239 & 0.028 & \citealt{GRACZ-SBCR} \\
\cline{2-6}
& 5.228 & 0.027 & 5.211 & 0.028 & \citealt{SALSI} \\
\cline{2-6}
& 5.281 & 0.027 & 5.3 & 0.027 & \citealt{SBCR-B2025} \\
\hline
\hline
\multirow{5}{*}{V83}& 5.024 & 0.019 & 4.925 & 0.018 & \citealt{KER_SBCRI} \\
\cline{2-6}
& 5.112 & 0.019 & 5.035 & 0.019 & \citealt{KER_SBCRII} \\
\cline{2-6}
& 4.905 & 0.024 & 4.914 & 0.024 & \citealt{GRACZ-SBCR} \\
\cline{2-6}
& 4.896 & 0.023 & 4.877 & 0.023 & \citealt{SALSI} \\
\cline{2-6}
& 4.969 & 0.024 & 4.982 & 0.024 & \citealt{SBCR-B2025} \\
\hline
\hline
\end{longtable}
\end{center}

\end{appendix}
\end{document}